\documentclass[12pt,a4paper]{article}

\usepackage[utf8]{inputenc}
\usepackage[T1]{fontenc}
\usepackage{lmodern}
\usepackage{microtype}
\usepackage[a4paper,margin=2.5cm]{geometry}

\usepackage{amsmath, amssymb}

\usepackage{graphicx}
\usepackage{subcaption}
\usepackage{svg}
\usepackage{tikz}
\usepackage{pdfpages}
\usepackage{wrapfig}

\usepackage{booktabs}
\usepackage{longtable}
\usepackage{array}
\usepackage{multirow}
\usepackage{adjustbox}
\usepackage{threeparttable} 
\usepackage{siunitx}       
\usepackage{lscape}

\usepackage{float}
\usepackage{xcolor}
\usepackage{soul}
\usepackage{caption}
\usepackage{authblk}

\usepackage{natbib}

\usepackage{hyperref}
\hypersetup{
    colorlinks=true,
    linkcolor=blue,
    citecolor=blue,
    urlcolor=cyan
}

\newcommand{\makeDashboard}[2]{
    \begin{figure}[H]
        \centering
        \begin{subfigure}[b]{0.32\textwidth}
            \centering
            \includegraphics[width=\linewidth]{SI_img/#1_Res_#2.pdf}
            \caption{Residuals}
        \end{subfigure}
        \hfill
        \begin{subfigure}[b]{0.32\textwidth}
            \centering
            \includegraphics[width=\linewidth]{SI_img/#1_ACF_#2.pdf}
            \caption{ACF}
        \end{subfigure}
        \hfill
        \begin{subfigure}[b]{0.32\textwidth}
            \centering
            \includegraphics[width=\linewidth]{SI_img/#1_PACF_#2.pdf}
            \caption{PACF}
        \end{subfigure}
        \caption{Diagnostic plots for \textbf{#1} (#2).}
    \end{figure}
}

\title{\textbf{State-dependent marginal emission factors with autoregressive components}}

\author[1,2]{Antonio Panico\thanks{Email: antonio.panico@unipr.it}}
\author[2]{Andrew Burlinson}
\author[1]{Luigi Grossi}

\affil[1]{Department of Engineering for Industrial Systems and Technologies, University of Parma, 43124 Parma, Italy}
\affil[2]{School of Economics, University of Sheffield, Sheffield, UK}

\date{} 

\begin{document}

\maketitle

\begin{abstract}
Accurate estimation of Marginal Emission Factors (MEFs) is critical for evaluating the decarbonization
potential of low-carbon technologies and demand-side management. However, canonical methodologies,
predominantly relying on linear regression and differencing techniques, fail to capture the structural non-linearities
inherent in the merit order, i.e. the marginal technology setting electricity prices. Utilizing Markov switching autoregressive models with exogenous
regressors (MS-ARX) and hourly US data (2019–2025), we identify distinct, mutually exclusive regimes governed
by fuel-price dynamics. We find that linear models overestimate abatement potential by masking the
dichotomy between a gas-driven and coal-driven marginal system. Furthermore, using robust structural
break detection, we link regime instability to a specific structural shift in natural gas pricing in May 2022. Our
results indicate that post-2022, the grid has transitioned into a correction phase where the coal-driven regime is less
persistent but highly volatile, necessitating state-dependent policy metrics rather than static annual averages.
\end{abstract}

\noindent \textbf{Keywords:} climate policy, electricity, emissions, regime-switching models, structural break \\
\noindent \textbf{JEL codes:} C22, C51, Q41, Q53 \\[5pt]
\section{Introduction}

The United States (U.S.) officially withdrew from the Paris Agreement on January 27, 2026. Marking the second time the nation exited the landmark climate pact under a Trump administration,  reaffirming its intent to renege on international legal obligations -- including the pledge to halve greenhouse gas emissions by 2030  (\citealp{usa2021Biden}). Despite the \cite{doe2023} arguing that achieving 100$\%$ clean electricity by 2035 could reduce energy-related emissions by 2.4 gigatons, the path toward deep decarbonization faces renewed uncertainty. Furthermore, established methods overlook inherent non-linearities in the displacement of fossil-fuels at the margin. This is problematic as recent executive orders actively target the revitalization of coal-based power. Our study addresses this critical gap in the literature by providing a robust econometric framework to capture these evolving marginal dynamics.

Policy evaluation often relies on fixed average emission factors (AEF), such as those produced by the \cite{USEPA2019}. Yet, it has become clear that this approach leads to significant (downward) bias, as renewable generation displaces the \textit{marginal} generator rather than the average fuel mix. Accurate estimation of Marginal Emission Factors (MEFs) is therefore essential for evaluating the decarbonization potential of renewable energy, electric vehicles, and demand-side management (\citealp{GraffZivin2014}). Indeed,
\cite{Holland2022} highlighted this tension clearly by showing how MEFs were on the rise in the U.S. despite falling average emissions at least in the short-run. While \cite{Gagnon2022} note the importance of structural change in long-run planning, short-run MEFs remain indispensable for context-dependent near-term operational policies and real-world behavioral responses (\citealp{Holland_Reply}).

The literature has developed several statistical approaches to estimate MEFs. \citet{Holland2008} pioneered a linear regression approach using U.S. data, regressing emissions on generation while controlling for temporal and regional fixed effects. This methodology was widely adopted, most notably by \citet{ Callaway2018}, \citet{Holladay2017}, and \citet{Holland2022}. In contrast, \citet{Hawkes2010} utilized differencing to represent inter-temporal changes, and \citet{Beltrami2020} utilized time series methods to model autocorrelation in the data-generating process. Although informative, these approaches share a common limitation: they assume a stable, linear relationship between generation and emissions. Moreover, linear techniques estimate an \textit{average} marginal effect. We argue that this fails to capture the inherent non-linearities in merit-order dispatch, where the marginal fossil fuel shifts between ``clean'' (e.g., natural gas) and ``dirty'' (e.g., coal) technologies depending on system load and prices.

We address this gap in the literature by utilizing Markov switching autoregressive models with exogenous regressors (MS-ARX)\footnote{Throughout this work, MS-ARX and MSM are used interchangeably to refer to the same model.} and hourly data for the lower 48 United States (US48). Our work complements relevant literature in which Markov-switching models (MSMs) have been applied in order to model power price forecasts (\citealp{KosterMosler2006} and \citealp{Kapoor2023}) and wholesale price formation (\citealp{Zachmann2007}). More recently, \citet{Holladay2017} used MSMs to identify structural breaks in the emissions-gas price relationship following the fracking boom. 

More closely related to our study is that of \citet{BenAmor2025} who suggest that regime-switching frameworks outperform dynamic linear regressions for MEF estimation in the German market. Their Markov switching dynamic regression (MSDR) model finds three regimes representing i) low-emission periods with renewable dominance, ii) high-emission periods characterized by fossil fuel dispatch, and iii) volatile transition states. However, despite finding the emissions and generation series to be stationary, the authors use first-order differences. This transformation is statistically unnecessary for stationary data and introduces interpretational ambiguity by modeling rates of change rather than level-based physical realizations. Another notable limitation of their approach is the use of total generation (including renewables) as the explanatory variable. Since renewable generation contributes zero direct emissions, their regime-switching mechanism distinguishes between periods of high renewable penetration (low MEF) and periods of fossil fuel dominance (high MEF), rather than capturing the shifting marginal dynamics between gas and coal technologies.

Our work overcomes such limitations by simultaneously grounding regime identification in fuel market fundamentals; specifically, we validate that observed regimes correspond to distinct marginal fossil-fuel states\footnote{While the emission intensity of oil is comparable to coal, we exclude oil here. Oil-fired units function almost exclusively as super-peakers, meaning they are invariably the marginal fuel when active. The modeling challenge and focus of this study is distinguishing between gas and coal at the margin.} and link regime instability to structural breaks in natural gas pricing. Moreover, our study shows that earlier linear models, despite being less rigorous for estimating MEFs, remain useful for analysing other behaviours of marginal units, specifically the \textit{marginal fuel probability} and \textit{generation--load responsiveness}.

We contribute to the literature in three main ways. First, by  rigorous estimation of MEFs for the total US48 region using an MS-ARX framework, allowing us to disentangle distinct, mutually exclusive operating regimes. Second, we demonstrate that standard linear models mask the dichotomy between ``clean'' (gas-margin) and ``dirty'' (coal-margin) fossil-fuel states, leading to potential overestimation of abatement potential. Third, we employ a robust structural break detection on natural gas prices to further unpack our regime-switching results. We find that a structural shift in gas pricing in May 2022 fundamentally altered the grid's stability. Specifically, we observe that after an inflationary phase characterized by energy volatility and rising prices driven by the COVID-19 pandemic and the Ukraine-Russia war, the energy market then entered a correction phase.

The rest of the paper is organized as follows. Section \ref{sec: Review} reviews existing MEF estimation methods. Section \ref{sec:Data description} presents the data, while the empirical methodology is described in Section \ref{sec: model spec}. The results are discussed in  Section \ref{sec: res and disc} and Section \ref{sec: Conclusion} offers conclusions.

\section{Review of existing methods}
\label{sec: Review}
The extant literature on 
MEF estimation follows three primary methodological frameworks. First, \textit{US fixed-effects} (US-FE), originating from  regression analysis of U.S. inter- and intra-day emissions and generation data while controlling for time and regional fixed effects (\citealp{Holland2008}, and \citealp{Callaway2018}). Second, the \citet{Hawkes2010} approach, which employs first-order differencing, capturing system-level changes between settlement periods i.e. intra-day. Third, \textit{ARIMA-FE}, introduced by \citet{Beltrami2020}, incorporates autoregressive integrated moving average (ARIMA) components alongside fixed effects within and across days. We will focus on intra-day in what follows, as this underpins our main results, noting that the arguments apply to inter-day, too.

The general \textit{US-FE} specification can be defined as:

\begin{equation}
E_{rt} = \beta_{r}G_{rt} + \alpha_{r} + \varepsilon_{rt}
\label{eq:usfe}
\end{equation}

\noindent where $E_{rt}$ represents emissions in region $r$ at time $t$, $G_{rt}$ denotes electricity generation, $\alpha_{r}$ captures regional fixed effects, and $\beta_{r}$ is the regional MEF. Since \citet{Holland2008} evaluated the environmental impacts of regional real-time pricing, similar specifications have been applied to renewable generation deployment (\citealp{Callaway2018}), electricity storage (\citealp{Carson2013}), electric vehicles (\citealp{GraffZivin2014}), and natural gas price shocks related to fracking (\citealp{Holladay2017}). However, this approach assumes the error term $\varepsilon_{rt}$ is independently distributed after controlling for fixed effects. If emissions exhibit autoregressive or moving average dynamics, the model may suffer from omitted variable bias and inefficient parameter estimates.

\citet{Hawkes2010} proposed an alternative framework based on first-differencing:

\begin{equation}
\Delta E_{t} = \beta\Delta G_{t} + \varepsilon_{t},
\label{eq:hawkes}
\end{equation}

\noindent where $\Delta E_{t} = E_{t} - E_{t-1}$ measures the change in emissions between two consecutive hours, and $\Delta G_{t} = G_{t} - G_{t-1}$ denotes the corresponding change in electricity generation. Using Great Britain (GB) data (2002-2009), \citet{Hawkes2010} estimated MEFs of approximately 690 kg CO$_2$/MWh, substantially exceeding the system AEF of 510 kg CO$_2$/MWh. Hawkes proposed differencing to interpret changes in the variables while, inadvertently, addressing the issue of non-stationary processes. Yet, as shown in \citet{Beltrami2020}, standard tests most often fail to reject the hypothesis of stationarity. The evidence presented herein for the U.S., mirroring that of Europe (\citealp{Beltrami2020}, and \citealp{BenAmor2025}), finds the relevant series to exhibit stationary properties. Therefore, Hawkes' model does not represent, from a statistical perspective, the most parsimonious approach for estimating intra-day MEFs. 

Indeed, \citet{Beltrami2020} shows that emissions and generation series for GB and Italy are stationary ($d=0$). Consequently, their empirical specification simplifies to:

\begin{equation}
\Phi_p(L)(E_{t} - \beta G_{t}) = \alpha + \Theta_q(L)\varepsilon_{t}
\label{eq:arma}
\end{equation}

\noindent with $p$ and $q$ selected via Akaike Information Criterion (AIC) and Bayesian Information Criterion (BIC), typically yielding ARIMA$(1,0,1)$, 
ARIMA$(2,0,1)$, or ARIMA$(1,0,2)$ specifications. The authors demonstrate that the \textit{ARIMA-FE} model consistently outperform both \textit{US-FE} and \textit{Hawkes} approaches according to information criteria. Crucially, their time series models generate more stable MEF trajectories with lower variance and coefficients of variation. 

However, all three approaches rely on linear models that yield a single average marginal effect, invariant to the state of the system. We argue that this is ill-suited to the inherently non-linear nature of emissions and power generation.
This motivates our use of MSMs, which allow the marginal emission factor to vary across latent regimes, thereby capturing the structural non-linearities that linear methods cannot accommodate.

\section{Data sources and description}
\label{sec:Data description}
\begin{figure}[h]
    \centering
    \includegraphics[width=1\linewidth]{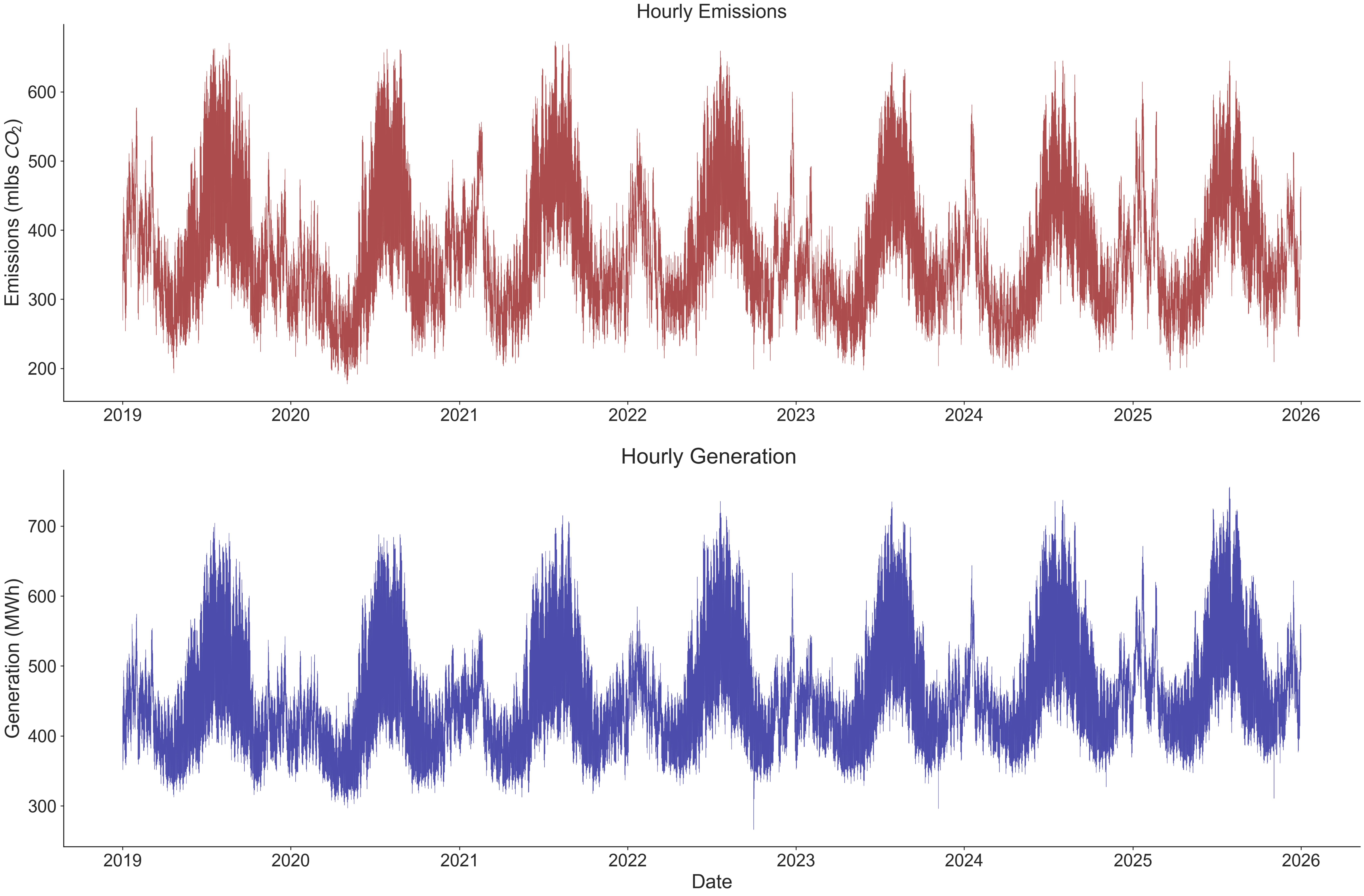}
    \caption{\footnotesize \textbf{Emissions and Generation:} Emissions are reported in $\text{Mlbs CO}_2$, while Generation in $\text{MWh}$ refers to the total fossil fuel generation.}
    \label{fig:hourly_gen}
\end{figure}

We construct our dataset using hourly electricity generation and emissions series for the lower 48 United States, obtained from  \citet{EIA2026} (EIA), see Figure \ref{fig:hourly_gen}. The sample covers the period from 1 January 2019 to 31 December 2025, with a complete series of $T=61,368$ hourly observations.

In line with relevant literature (\citealp{Callaway2018}, and \citealp{Holland2022}), we rely on actual hourly generation rather than system load. Indeed, while load data is often used to approximate demand, it fails to account for the carbon intensity of imported electricity, e.g., cross-border flows from Canada or Mexico.

The aggregate CO$_2$ emissions $E_t$ for each hour $t$ can be represented as:
\begin{equation}
    E_t = \sum_{f} G_{f,t} \cdot \theta_f
    \label{eq:emissions_calc}
\end{equation}
where $G_{f,t}$ denotes the hourly generation of fuel type $f$ (i.e., coal or natural gas) and $\theta_f$ represents the respective emission factor. 

Table \ref{tab:overall_summary} presents the summary statistics for the full sample. The U.S. power system generated 461,092 MWh per hour, on average, and mean hourly emissions of approximately 168,050 lbs of CO$_2$. The high standard deviation in generation reflects the significant temporal heterogeneity of the grid.

\begin{table}[ht]
\centering
\caption{\footnotesize \textbf{ Summary Statistics (2019--2025).}  }
\label{tab:overall_summary}
\small
\begin{tabular}{lccccc}
\hline
\textbf{Variable} & \textbf{N} & \textbf{Mean} & \textbf{Std. Dev.} & \textbf{Min} & \textbf{Max} \\
\hline
Emissions (lbs) & 61,368 & 168,050 & 41,036 & 80,496 & 305,224 \\
Generation (MWh) & 61,368 & 461,092 & 76,092 & 266,277 & 755,638 \\
\hline
\end{tabular}
\end{table}

The dynamics are detailed in Table S1 of the Online Supplement (OS). Unlike many European systems that peak in winter (\citealp{Beltrami2022}), the U.S. is characterized by a dominant summer peak, driven by cooling demand. During summer months, mean generation surges to nearly 550,000 MWh, frequently forcing the dispatch of carbon-intensive peaker plants, which drives emissions to their annual maximums of about 219,000 lbs. The system reaches a trough during spring, where mild load and high renewable penetration reduce mean emissions to their minimum of 124,000 lbs.

\subsection{Stochastic properties and empirical justification for Markov Switching Models}

In this section, we examine the stochastic properties of CO$_2$ emissions to determine the appropriate modeling framework. Specifically, we verify whether the data, after controlling for deterministic seasonality, exhibits stationarity while simultaneously possessing nonlinear characteristics.

To isolate the stochastic component, we first extracted the residuals from a linear regression that filters out deterministic seasonal effects using hourly, daily, and monthly indicators and a linear time trend. We then subjected these filtered residuals to a comprehensive battery of diagnostic tests, the results of which are summarized in Table \ref{tab:diagnostics_summary}.

\subsubsection*{Stationarity analysis}
Table \ref{tab:diagnostics_summary} (Panel A) reports the results of three complementary tests used to assess the presence of unit root and stationarity: the Augmented Dickey-Fuller (ADF), Phillips-Perron (PP), and Kwiatkowski-Phillips-Schmidt-Shin (KPSS) tests.

The lag lengths for the ADF test were selected via the BIC. The criterion consistently identified a range of 24--27 lags, capturing the physical 24-hour diurnal cycle of the power system. For the PP and KPSS tests, the spectral bandwidth was determined using the Newey-West estimator. We found a bandwidth between 37--55, ensuring robustness against the serial correlation and time-varying volatility inherent in high-frequency data.

The results show strong evidence of stationarity (see Panel A). Both the ADF and PP tests reject the null hypothesis of a unit root at the 1\% significance level for every year in the sample, while the KPSS test fails to reject the null hypothesis of stationarity.

\subsubsection*{Nonlinearity and complexity}
Having established stationarity, we investigated the residuals for nonlinear dependencies that standard linear models cannot capture. Panel B of Table \ref{tab:diagnostics_summary} presents the results of Tsay’s F-test for nonlinearity, the BDS test for independence, and the Peña-Rodríguez (PR) test for volatility clustering.

To select the lag order $p$, we fitted a sequence of AR($p$) models to the residuals and selected the optimal order that minimized the AIC. We consistently found an optimal lag of $p=11$ or $12$ hours, reflecting the semi-diurnal operational cycles of the power system. Using this optimal lag, Tsay’s F-test strongly rejects the null of linearity ($p < 0.01$) for all years, indicating the presence of significant quadratic interactions. Furthermore, the BDS test statistics clearly rejects the hypothesis of independent and identically distributed (i.i.d.) noise. Crucially, the PR test strongly rejects the null of no autoregressive conditional heteroscedasticity (ARCH) effects, confirming the presence of volatility clustering. 

These findings suggest that the underlying data generating process undergoes structural changes or regime shifts. This provides a robust empirical justification for adopting MSMs, which explicitly capture such regime-dependent dynamics.

\begin{table}[ht]
\centering
\footnotesize
\footnotesize
\setlength{\tabcolsep}{4pt}
\caption{ \footnotesize Diagnostic Tests for stationarity and nonlinearity. 
This table reports the test statistics for the  emission residuals. 
\textbf{Panel A} reports stationarity tests: ADF, PP, and KPSS. 
\textbf{Panel B} reports nonlinearity tests: Tsay's F-test, BDS test ($m=6$), and PR test for volatility clustering.}
\begin{tabular}{lcccccc}
\hline

& \multicolumn{3}{c}{\textbf{Panel A: Stationarity}} & \multicolumn{3}{c}{\textbf{Panel B: Nonlinearity}} \\
\hline
\textbf{Year} & \textbf{ADF} & \textbf{PP} & \textbf{KPSS} & \textbf{Tsay's F} & \textbf{BDS} & \textbf{PR} \\
\hline

2019 & -8.22$^{***}$ & -8.66$^{***}$ & 0.058 & 2.68$^{***}$ & 198.72$^{***}$ & 13.70$^{***}$ \\
2020 & -8.79$^{***}$ & -9.71$^{***}$ & 0.059 & 2.01$^{***}$ & 203.59$^{***}$ & 13.48$^{***}$ \\
2021 & -8.24$^{***}$ & -8.59$^{***}$ & 0.046 & 2.33$^{***}$ & 219.82$^{***}$ & 13.77$^{***}$ \\
2022 & -8.95$^{***}$ & -8.45$^{***}$ & 0.050 & 3.40$^{***}$ & 220.18$^{***}$ & 13.71$^{***}$ \\
2023 & -8.50$^{***}$ & -9.40$^{***}$ & 0.045 & 1.81$^{***}$ & 197.40$^{***}$ & 13.61$^{***}$ \\
2024 & -8.02$^{***}$ & -8.69$^{***}$ & 0.046 & 1.47$^{***}$ & 203.75$^{***}$ & 13.79$^{***}$ \\
2025 & -7.70$^{***}$ & -8.23$^{***}$ & 0.044 & 1.77$^{***}$ & 197.91$^{***}$ & 14.88$^{***}$ \\
\hline
\\
\multicolumn{7}{l}{\footnotesize $^{***}$, $^{**}$, and $^{*}$ indicate significance at the 1\%, 5\%, and 10\% levels, respectively.} \\
\multicolumn{7}{l}{\footnotesize Note: For KPSS, a low statistic (no stars) indicates failure to reject stationarity.} \\
\end{tabular}

\label{tab:diagnostics_summary}
\end{table}

\section{Model specification and estimation methodology}
\label{sec: model spec}

We capture the dynamics of fossil fuel generation operating at the margin using a two-state MS-ARX. In this section, we define the model specification and detail the estimation procedure, which utilizes the expectation-maximization (EM) algorithm proposed by \citet{dempster1977} and adapted for time series by \citet{hamilton1990}. This iterative procedure integrates the  \citep{Hamilton1989} filter for forward inference and the \citep{kim1994} smoother for backward inference to facilitate maximum likelihood estimation (MLE).

\subsection{Model specification}

Let $y_t$ denote the observed emission level at time $t$ for $t = 1, \ldots, T$, and $x_{1,t}$, $x_{2,t}$ for $t = 1, \ldots, T$ the renewable and non-renewable\footnote{In practice, we estimate non renewable generation as the sum of hourly generation from coal, gas provided in \cite{EIA2026}.} 
 generation covarites.
The emissions process is governed by an unobserved discrete state variable $S_t \in \{1, 2\}$\footnote{Although selection criteria suggest three states, an inspection of smoothed probabilities showed a third regime was redundant; thus, a two-state model was selected (see Section S2 of the Online Supplement for further details).}. We interpret these regimes based on the fuel mix operating at the margin, namely we interpret a \textit{Low-Emission state} if marginal generation is dominated by efficient natural gas units and \textit{High-Emission state} if marginal generation shifts toward carbon-intensive coal or low-efficiency peaking units.

Conditional on the regime $S_t$, the observation equation is:
\begin{equation}
  y_t = \alpha_{S_t} + \phi_{S_t} y_{t-1} + \beta_{1,S_t} x_{1,t} + \beta_{2,S_t} x_{2,t} + \epsilon_t, \quad \epsilon_t \sim \mathcal{N}(0, \sigma^2)
 \label{eq:observation}
\end{equation}
where $\phi_{S_t}$ is the autoregressive term, $\beta_{1,S_t}$ is the \textit{regime-specific} MEF, representing the marginal intensity of the grid in state $S_t$ and  $\beta_{2,S_t}$ is the marginal probability of renewable sources in state $S_t$. The state variable $S_t$ evolves according to a first-order ergodic Markov chain with transition matrix $\mathbf{P}$, where $p_{ij} = \mathbb{P}(S_t = j \mid S_{t-1} = i)$.

Although more general MSMs allowing for regime-dependent heteroscedasticity are widely used in the literature (\citealp{Bauwens2010}, and \citealp{Kang2014}), we impose a constant variance across regimes ($\sigma_1^2 = \sigma_2^2 = \sigma^2$) given theoretical and empirical considerations. Theoretically, this restriction ensures that regime identification is driven by structural shifts in merit order dispatch rather than transient volatility clustering. 
Indeed, regime‑specific variances leads to weak identification of the mean parameters, as the likelihood tends to classify observations by volatility rather than by differences in marginal emissions (\citealp{hs1994}, and \citealp{swz2008}). Empirically, we find that heteroscedastic Markov switching  specifications systematically produce negative marginal emission factors in the low-emission regime, a physical impossibility that indicates model misspecification.
This behavior arises because heteroscedastic MSMs tend to classify regimes primarily by variance rather than mean structure. In the electricity context, high-volatility periods (e.g., peak demand hours with rapid unit cycling) may coincide with either high or low marginal emissions depending on the fuel mix. When the model partitions observations by volatility, the resulting mean parameters become biased, yielding negative coefficients that are empirically inconsistent with MEFs.

\subsection{Likelihood and estimation strategy}

The estimation of the parameter vector $\theta = (\boldsymbol{\beta}_1, \boldsymbol{\beta}_2, \sigma^2, p_{11}, p_{22})$ relies on MLE. However, because the state sequence $\mathcal{S} = (S_1, \dots, S_T)$ is unobserved, we cannot maximize the complete-data likelihood directly.

The quantity of interest is the \textit{observed-data log-likelihood}, which marginalizes over all possible state sequences. Let $\mathcal{Y}_T$ denote the full history of observations. The observed log-likelihood is:
\begin{equation}
    \mathcal{L}_{obs}(\theta) = \sum_{t=1}^T \log f(y_t \mid \mathcal{Y}_{t-1}; \theta).
    \label{eq:obs_likelihood}
\end{equation}
Direct maximization of \eqref{eq:obs_likelihood} is computationally intractable because the marginal density $f(y_t \mid \dots)$ involves a summation over $2^T$ possible state paths.

To overcome this \citet{hamilton1990} developed a procedure which relies on the EM algorithm proposed by \citet{dempster1977}.  Let us assume that a complete realization of the sequence $\mathcal{S} = (S_1, \dots, S_T)$ is known , it is possible to define the complete-data log-Likelihood  as :
\begin{equation}
    \mathcal{L}_c(\boldsymbol{\theta}; \boldsymbol{y}, \mathcal{S}) = \sum_{t=1}^T \log f(y_t \mid S_t, \mathbf{x}_t; \boldsymbol{\theta}) + \sum_{t=1}^T \log p_{S_{t-1}S_t}.
\end{equation}

However, in practice the sequence $\mathcal{S} = (S_1, \dots, S_T)$ is unknown, so the role of the EM algorithm is to iteratively maximize the expected value of the complete-data log-likelihood (the Q-function) conditional on the observed data and the parameters from the previous iteration, $\theta^{(v)}$. In other words, the EM algorithm is divided into two steps: the (i) E-Step (Expectation) and the (ii) M-Step (Maximization).

The E-Step is where we compute the expectation of the log function given   $\theta^{(v)}$ and y:
    \begin{equation}
    Q(\boldsymbol{\theta} \mid \boldsymbol{\theta}^{(v)}) = \mathbb{E}_{\mathcal{S} \mid \boldsymbol{y}, \boldsymbol{\theta}^{(v)}} \left[ \log L_c(\boldsymbol{\theta}; \boldsymbol{y}, \mathcal{S}) \right].
\end{equation}
This step requires computing the smoothed posterior probabilities of the regimes:
\begin{align}
    \xi_{t|T}^{(k)} &= \mathbb{P}(S_t = k \mid \boldsymbol{\mathcal{Y}}_T; \boldsymbol{\theta}^{(v)}) \quad k = 1,2\quad \text{(Smoothed Probabilities)}, \\
    \xi_{t-1,t|T}^{(ij)} &= \mathbb{P}(S_{t-1}=i, S_t=j \mid \boldsymbol{\mathcal{Y}}_T; \boldsymbol{\theta}^{(v)}) \quad \text{(Smoothed Joint Probabilities)}.
\end{align}
The M-Step updates the parameters by finding the value that maximizes the expected complete-data log-likelihood:
\begin{equation}
    \boldsymbol{\theta}^{(v+1)} = \underset{\boldsymbol{\theta}}{\arg\max} \, Q(\boldsymbol{\theta} \mid \boldsymbol{\theta}^{(v)}).
    \label{eq:m_step_argmax}
\end{equation}

These steps are repeated until the increase in the observed likelihood \eqref{eq:obs_likelihood} falls below a convergence tolerance. It has been shown that the EM algorithm converges to a local maximum \citep{dempster1977}.\footnote{For further details on the Maximization step we refer the reader to \cite{hamilton1990}.}

\section{Results and discussion}
\label{sec: res and disc}

In this section, we present the findings of our Markov-Switching analysis applied to the US48 hourly emissions and generation data. We first validate the model specification, distinguishing our regime-switching approach from canonical linear benchmarks. Subsequently, we analyze the characteristics of the identified regimes and link their stability to the structural dynamics of the natural gas market.

\subsection{Model comparison}

The primary advantage of the MS-ARX framework over classical approaches, discussed in Section \ref{sec: Review}, is its ability to decompose the emission data generating process into distinct, mutually exclusive regimes. This allows for a time-varying estimation of MEFs that reflects the changing marginal fuel mix (i.e., coal vs. gas) rather than a static average.

A critical modeling decision in regime-switching frameworks is the selection of the autoregressive order ($p$). While preliminary diagnostics on the de-seasonalized emission series using Partial Autocorrelation Functions (PACF)\footnote{ The plots of ACF and PACF for generation and emission are reported in Sections S3 and S4  of the Online Supplement} suggested potential higher-order dependence, we performed a sensitivity analysis which reveals that increasing the lag order yields no substantial improvement in model fit while introducing severe instability. As reported in Table S4 of OS, estimating MS-ARX models with $p \geq 2$ consistently failed to converge across multiple optimization algorithms\footnote{This convergence failure is well-documented in the regime-switching literature (\citealp{hamilton1990}, and \citealp{Krolzig1997}).}. Furthermore, the marginal gains in likelihood for these unstable specifications were negligible, confirming that the additional complexity does not translate into meaningful improvements in explanatory power. 

We therefore adopt a parsimonious MS-ARX(1) specification\footnote{Section S6 of the
Online Supplement provides the residual diagnostics for MS-ARX(1).}. Conceptually, this imposes a clear division of labor: the AR(1) term captures short-run hour-to-hour persistence inherent in grid operations (e.g., thermal unit ramping constraints), while the regime-switching component absorbs slower-moving structural shifts in merit order (i.e., transitions from gas-driven to coal-driven states). Unlike the ARIMA specification, which smoothes over volatility, the regime-switching framework explicitly
disentangles the distinct operating dynamics inherent to the merit-order dispatch. This structural identification allows quantification of how emissions vary across different fuel-margin states.

To assess the performance of our approach, we compared MS-ARX against the benchmarks defined in Section \ref{sec: Review}. To select the optimal orders of the ARIMA model, we apply an automated selection procedure based on MLE. The lag order p and the moving average order q are jointly chosen to maximize the log-likelihood of the model, while the integration order d = 0 because the series are stationary as shown in Section \ref{sec:Data description}, leading to the best-fitting ARIMA(3,0,4) model.

A detailed comparison of model fit across ARIMA, MS-ARX and US-FE is reported in Table S5 of OS. The results reveal a clear ranking where ARIMA achieves the highest performance in terms of log-likelihood and lowest values across AIC, BIC, and Hannan-Quinn information criterion (HQIC), making it the best-fitting model, while MS-ARX places second. Both models exhibit a substantial gap relative to US-FE, reflecting the well-known limitations of linear models in capturing the nonlinear and persistent dynamics of the emissions. While MS-ARX does not achieve the best statistical fit compared to ARIMA, its core value lies in the structural interpretation it provides. More precisely, MS-ARX captures state dependent behavior that neither ARIMA nor US-FE can offer, making it more suited to identifying different regimes of emissions driven by the merit-order dispatch.

\begin{figure}[h]
\centering
\includegraphics[width=1\linewidth]{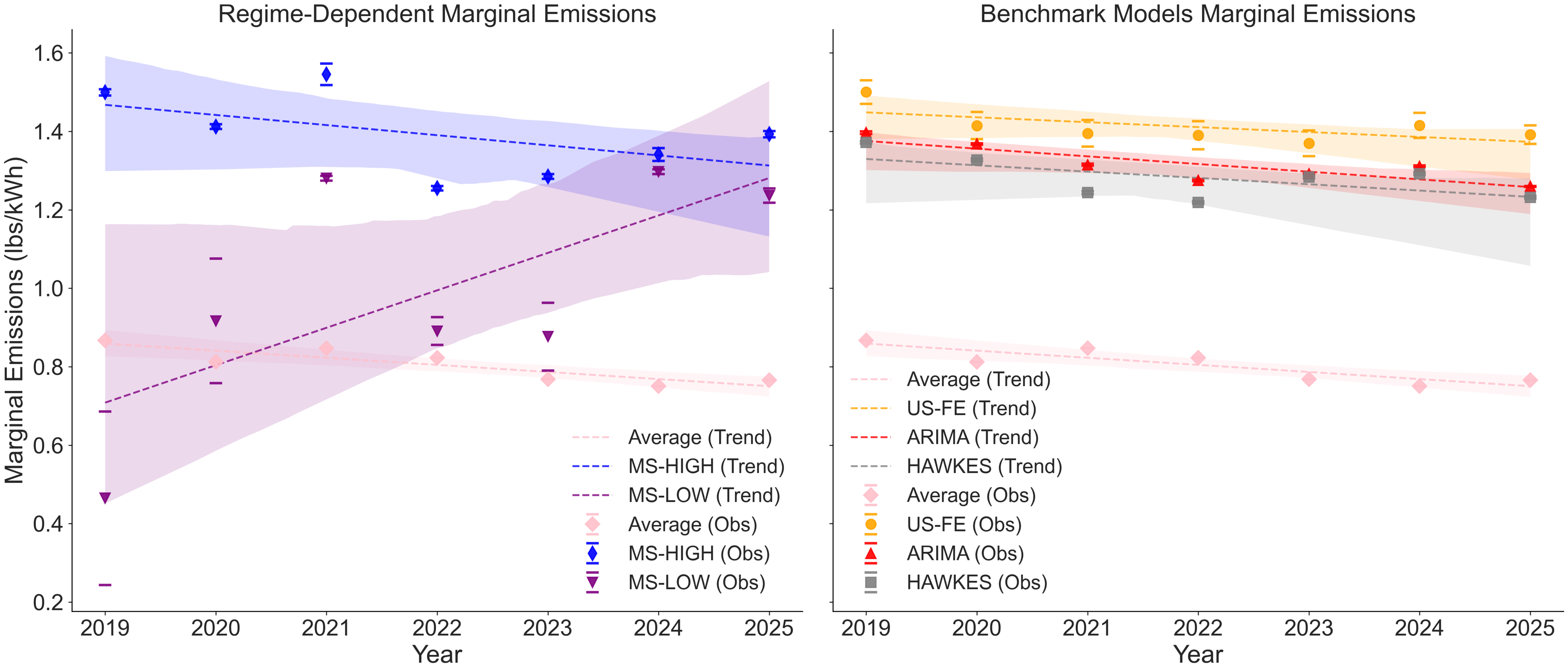}
\caption{\footnotesize \textbf{Temporal evolution of MEFs (2019--2025).} 
\textbf{(Left) Regime-dependent dynamics:} The trajectories of the High-MEF and Low-MEF regimes. Note the convergence of the two states in the end of the considered period.
\textbf{(Right) Benchmark Comparison:} Contrast between dynamic (ARIMA) and static (US-FE, HAWKES) linear specifications against the regime-switching baseline.}
\label{fig:mef_comparison}
\end{figure}

As illustrated in Figure \ref{fig:mef_comparison}, the distinct gap between average emissions (pink line) and all other models reaffirms that average emissions consistently underestimate marginal displacement potential. Moreover, in stark contrast to \citet{Holland2022}, our results show that for all methods we are comparing, MEFs follow a clear downward trend. Consistent with the discussion in Section \ref{sec: Review}, ARIMA-FE (red) provides a stable middle ground positioned between the US-FE (orange) and Hawkes (gray) estimates.

The \textit{High-MEF} regime (blue) takes values close to those produced by the ARIMA, Hawkes, and US-FE estimates. This suggests that the this state captures the ``dirty" marginal high fossil-fuel regime (i.e, predominantly coal). On ther other hand, the \textit{Low-MEF} regime (purple) yields smaller coefficients than the \textit{High-MEF} regime, but exhibits greater volatility, particularly over the period 2019–2022. This divergence indicates that when the grid enters a ``clean" fossil-fuel state, the marginal emission factor decouples from the linear aggregate trend.

By the end of the period, the two states converge. This alignment can be explained by the significant increase in the gas volume, driven by declining gas prices. As gas became cheaper, volumes increased, ultimately pushing the system toward a single dominant marginal source, in this case, gas. Standard linear models  smooth over these episodes, masking the fact that during specific regimes, the marginal carbon benefit of an additional unit of generation is substantially lower than the annual average would suggest. Detailed coefficient estimates and standard errors for all models are reported in Table S6 of OS.

\begin{figure}[ht!]
    \centering
    \includegraphics[width=1\linewidth]{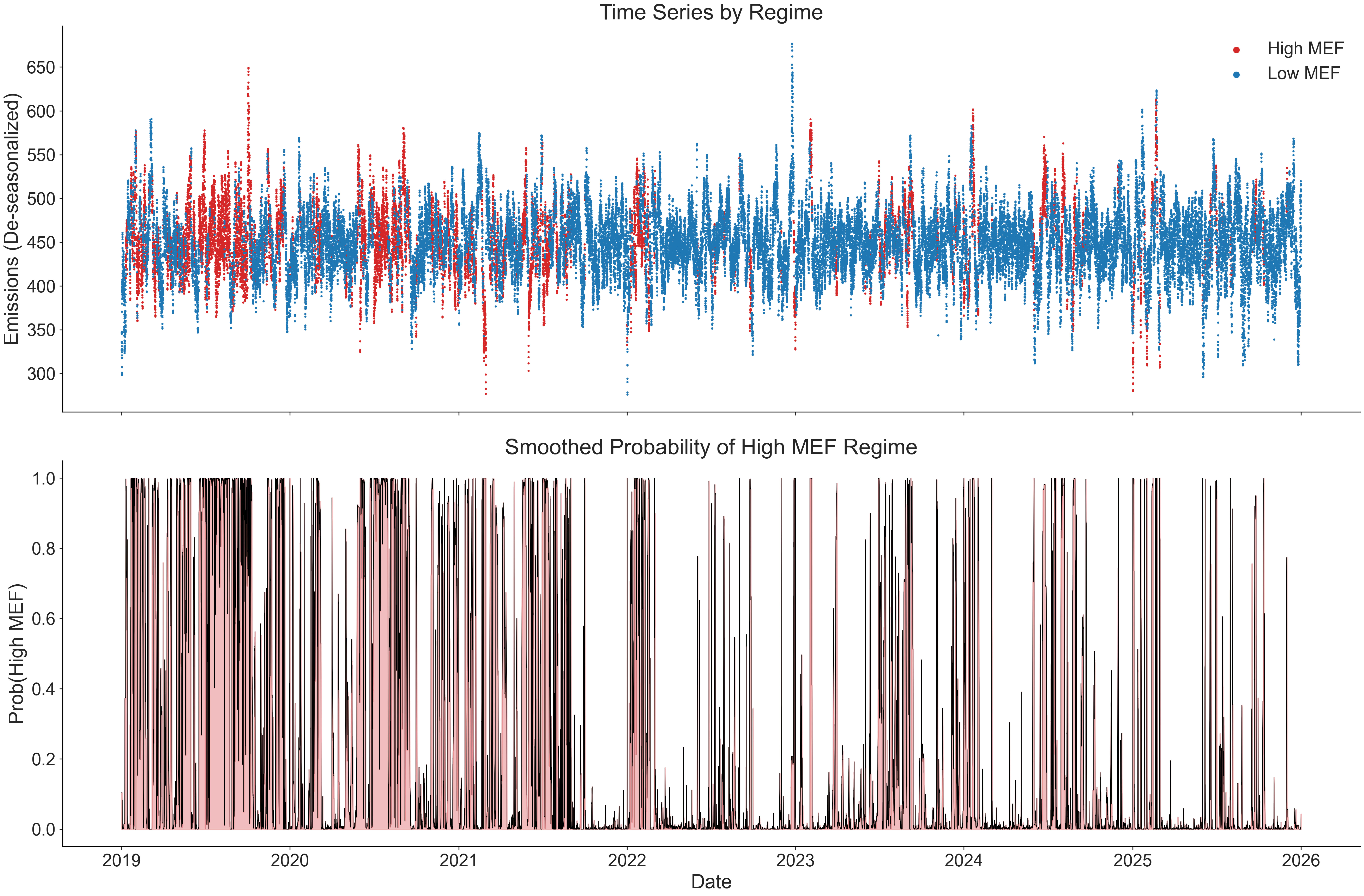} 
    \caption{\footnotesize \textbf{Regime Switching Dynamics (2019--2025).} Top: Hourly emissions colored by the most probable regime (Red = High MEF, Blue = Low MEF). Bottom: Smoothed probability $\mathbb{P}(S_t = \text{High MEF} | \mathcal{Y}_T)$. The distinct binary separation confirms that the grid operates in mutually exclusive states rather than a continuum.}
    \label{fig:regime_prob}
\end{figure}

\subsection{Temporal dynamics and drivers of regime switching}
\label{sec:regime_dynamics}

  MSM frameworks not only have the advantage of revealing temporal MEF dynamics but also are able to disentangle distinct mutually exclusive MEF states. Figure \ref{fig:regime_prob} illustrates this dichotomy, showing the time series of emissions by observed regime (top panel) and the smoothed probability of being in the \textit{High-MEF} state (bottom panel).

\subsubsection*{Volatility and structural shifts}
As evidenced by the smoothed probabilities, the \textit{High-MEF} and \textit{Low-MEF} regimes are rarely contemporaneous; the probability density is concentrated near 0 or 1, showing that at any given hour, the marginal technology is unambiguously defined. 

However, the persistence of these regimes changes substantially over time. Prior to 2022, the system exhibits prolonged periods of stability where the \textit{High-MEF} regime dominates. This behavior undergoes structural transformation coinciding with the last quarter of 2021, when the U.S. began experiencing 
heightened volatility in energy prices driven by a global energy crisis (we return to this in Section \ref{sec:structural_break}). Moreover, as shown in Figure \ref{fig:regime_prob} (bottom panel), the occurrence of the \textit{High-MEF} state becomes less persistent and more volatile, characterized by rapid switching rather than sustained blocks.

\subsubsection*{Seasonality of marginal emissions}
\begin{figure}[ht]
    \centering
    \includegraphics[width=1\linewidth]{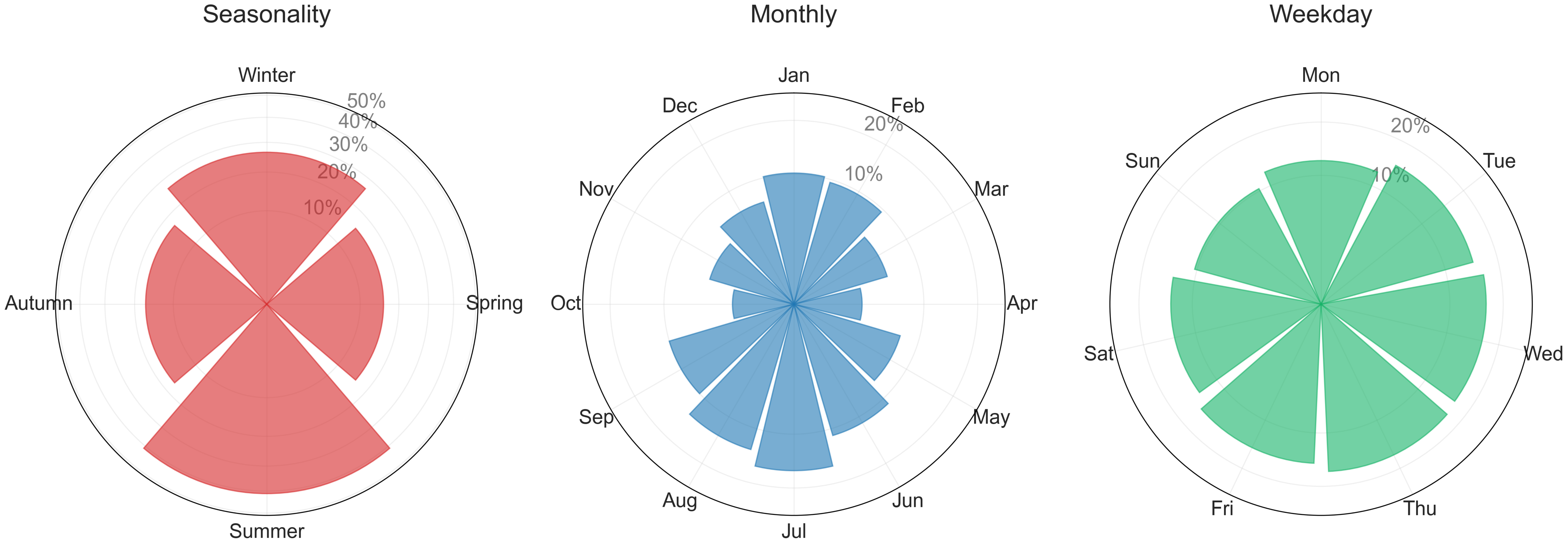} 
    \caption{\footnotesize \textbf{Temporal Distribution of the High-MEF Regime.} Radial plots showing the percentage occurrence of the High-Emission state by Season (Left), Month (Center), and Day of Week (Right).}
    \label{fig:regime_radar}
\end{figure}

To further investigate the drivers of these regimes, Figure \ref{fig:regime_radar} decomposes the occurrence of \textit{High-MEF} across seasonal and weekly cycles.

The \textit{High-MEF} regime is not randomly distributed, but follows a clear load-driven pattern. Seasonally, it peaks during periods of cooling (summer) and heating (winter). During these peaks less efficient, carbon-intensive coal or oil peakers are dispatched to meet the marginal demand, thereby triggering the high-emission state. Conversely, the \textit{High-MEF} probability contracts significantly during spring and autumn months, where mild demand allows cleaner gas units to set the price.

\subsubsection*{Identifying the marginal fuel}

To unpack the physical behavior of the regimes, we investigate the marginal response of specific fuel technologies to system load fluctuations. We employ a first-difference linear specification to isolate the ramping behavior of marginal units from baseload generation. For each fuel type $f$ and identified regime $k$, we estimate the following regression:

\begin{equation} \label{eq:marginal_fuel}
    \Delta G_{f,t} = \alpha_{f}^{(k)} + \beta_{f}^{(k)} \Delta Load_t + \epsilon_{f,t} \quad \text{for } S_t = k
\end{equation}

\noindent where $\Delta$ denotes the hourly change ($X_t - X_{t-1}$), $G_{f,t}$ is the generation of fuel $f$, and $Load_t$ represents the system demand.

The coefficient $\beta_{f}^{(k)}$ is defined as the \textit{marginal fuel probability} (\citealp{Holland2022}). It measures the proportion of an incremental megawatt (MW) of demand that is met by fuel $f$. Consequently, this value can be interpreted as the probability that a given fuel type is on the margin during regime $k$.

\begin{figure}[h]
    \centering
    \includegraphics[width=0.4\textwidth]{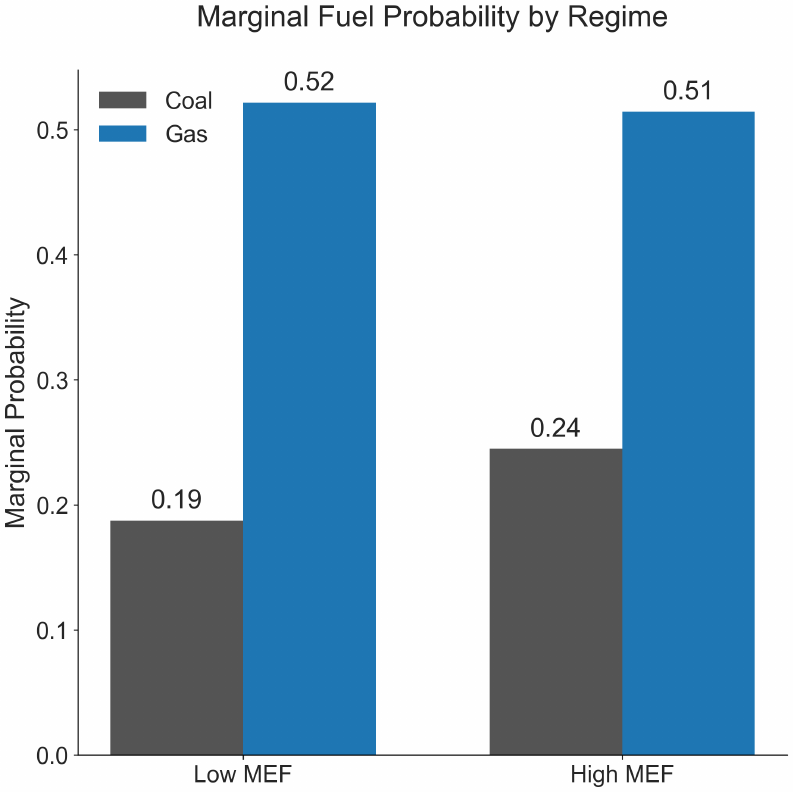}
    \caption{\footnotesize\textbf{Marginal Fuel Probability by Regime.} The bar chart displays the estimated $\beta$ coefficients from Equation (\ref{eq:marginal_fuel}). It illustrates that Gas is the dominant marginal fuel in the Low MEF regime, while Coal significantly increases its marginal contribution in the High MEF regime.}
    \label{fig:marginal_probability}
\end{figure}

Figure \ref{fig:marginal_probability} presents the estimated coefficients. There is a distinct shift in dispatch, whereby the \textit{High-MEF} regime exhibits a higher probability of coal acting as the marginal fuel compared to the low regime.

\subsection{Robust detection of structural break in natural gas markets}
\label{sec:structural_break}

To identify the exogenous drivers of the observed regime instability, we draw on the insights of \citet{Holladay2017} and \citet{knittel2015}. These studies demonstrate that structural breaks and changes in fuel prices can fundamentally alter MEFs by shifting the dispatch order of power plants. When natural gas is expensive, it is typically reserved for peaking capacity during high-demand periods, while relatively inexpensive coal serves as the baseload. However, as gas prices fall, efficient natural gas plants become cheaper to operate, displacing coal from the baseload and pushing it to the margin, where it sets a higher marginal emission rate for the system (\citealp{Holland2022}).

To analyze this shift, we examine daily Henry Hub natural gas spot prices\footnote{Data have been downloaded by \cite{EIA_HenryHub}.}. Indeed, natural gas prices exhibit marked volatility and frequent spikes (see Figure \ref{fig:gas_break}); this behavior is documented in the electricity market by \citet{GianfredaGrossi2012} and \citet{Grossi2019}. Although different procedures to detect instability are widely used in the electricity and financial markets (\citealp{apergis2015}, and \citealp{vriz2026}), many of them are highly sensitive to outliers and transient shocks. Given this susceptibility to extreme events, we adopt the robust monitoring approach proposed by \citet{riani2019robust} to detect a structural break in slope.

\subsubsection*{Detecting Structural break}
We model the daily natural gas price $P_t$ as a trend-stationary process subject to a potential regime change at an unknown time $\tau$. Specifically, we test for a structural break in the slope of the price trend, allowing us to distinguish between a stable long-term trajectory and the onset of a new pricing regime. The robust broken-trend model is specified as follows:

\begin{equation}
    P_t = \alpha + f(t) + 
    \delta (t - \tau) I(t > \tau) + \epsilon_t, \quad t = 1, \dots, T
    \label{eq:broken_trend}
\end{equation}

\noindent where $\alpha$ is the intercept, $f(t)$ represents the  non linear baseline price trend, and $\delta$ denotes the slope change that occurs after the break date $\tau$. The term $I(\cdot)$ is the indicator function, taking the value of 1 if $t > \tau$ and 0 otherwise. The error term $\epsilon_t$ is assumed to be independent, but may follow a heavy-tailed distribution contaminated by outliers.

\subsubsection*{Robust Estimation Strategy}
To estimate the unknown breakpoint $\tau$ robustly, we minimize the trimmed sum of squared residuals rather than the total sum of squares. Following \cite{riani2019robust}, we define the trimmed objective function $S(\tau)$ as:

\begin{equation}
    S(\tau) = \sum_{j=1}^{h} (r^2)_{(j)}(\tau)
    \label{eq:lts_objective}
\end{equation}

\noindent where $(r^2)_{(1)}(\tau) \leq (r^2)_{(2)}(\tau) \leq \dots \leq (r^2)_{(T)}(\tau)$ are the ordered squared residuals from the model estimated conditional on break date $\tau$. The trimming parameter $h$ represents the number of smaller squared residuals used for estimation. We set $h = 0.99T$, thereby effectively ignoring the 1\% most extreme observations during the search procedure to prevent masking effects.

The estimated break date $\hat{\tau}$ is the solution to the discrete optimization problem:

\begin{equation}
    \hat{\tau} = \underset{\tau \in \Gamma}{\arg \min} \ S(\tau)
    \label{eq:optimization}
\end{equation}

\noindent where $\Gamma = [\lfloor 0.15T \rfloor, \lfloor 0.85T \rfloor]$ is the search window, trimmed at the boundaries to ensure parameter stability (\citealp{andrews1993tests}).

\subsubsection*{Inference and Outlier Detection}
Since the asymptotic distribution of the raw LTS estimator is non-standard, we employ a reweighting procedure to perform valid statistical inference on the slope change parameter $\delta$ (\citealp{rousseeuw2018anomaly}).

First, we calculate a robust scale estimate $\tilde{\sigma}$ based on the minimized trimmed loss, adjusted for consistency with the normal distribution (\citealp{croux1992time}):

\begin{equation}
    \tilde{\sigma} = c_{h,T} \sqrt{\frac{S(\hat{\tau})}{h}}
\end{equation}

\noindent where $c_{h,T}$ is a consistency factor correcting for the truncation of the error distribution \footnote{Asymptotically, it is defined as $c_{h,T} = \sqrt{ \frac{1}{1 - \frac{2n}{h} q \phi(q)} }$, where $q = \Phi^{-1}\left(\frac{h+n}{2n}\right)$ is the quantile of the standard normal distribution corresponding to the trimming proportion, and $\phi(\cdot)$ and $\Phi(\cdot)$ denote the standard normal density and cumulative distribution functions, respectively (\citealp{rousseeuw1987robust}).}. Using this scale, we identify outliers by computing standardized residuals. An observation $P_t$ is assigned a weight $w_t$:

\begin{equation}
    w_t = 
    \begin{cases} 
    1 & \text{if } |r_t(\hat{\tau}) / \tilde{\sigma}| \leq \Phi^{-1}(0.995) \\
    0 & \text{otherwise}
    \end{cases}
    \label{eq:weights}
\end{equation}

Finally, we estimate the model in Equation (\ref{eq:broken_trend}) using weighted least squares. The statistical significance of the structural break is determined by the $t$-statistic of the estimate $\hat{t}_{\delta}$. We reject the null hypothesis of a stable trend if $|\hat{t}_{\delta}| > 1.96$, confirming a structural break candidate robust to distributional anomalies.

\begin{figure}[ht]
    \centering
    \includegraphics[width=1\linewidth]{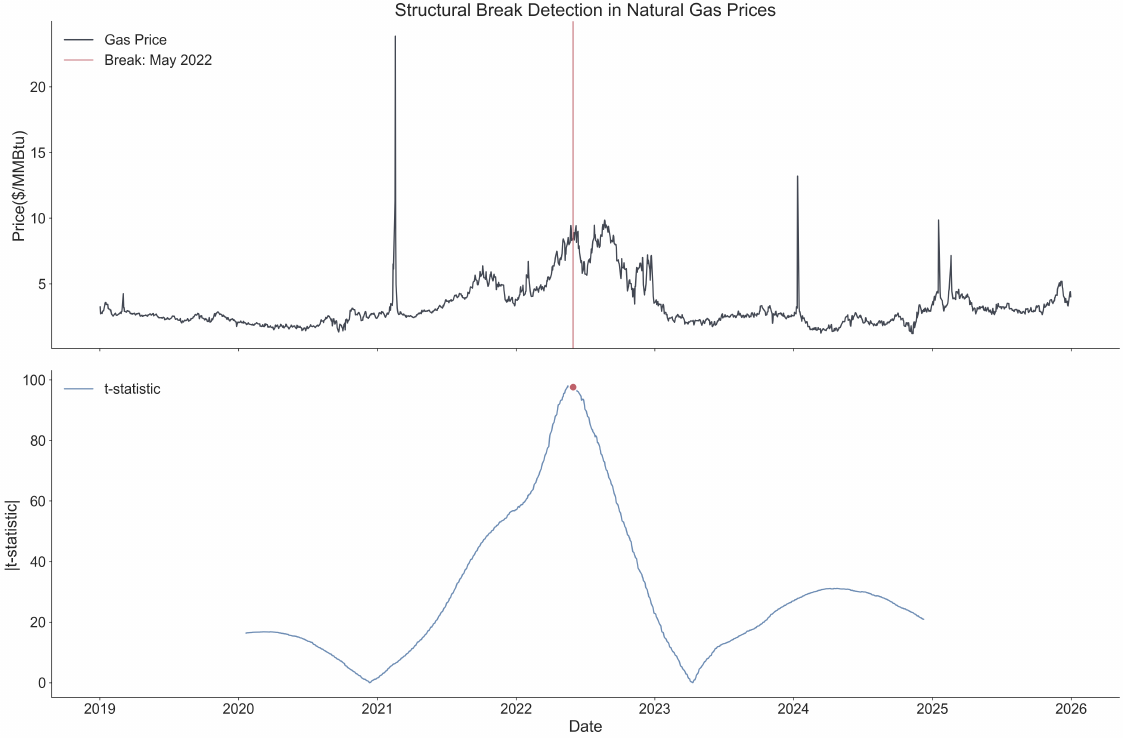} 
    \caption{\footnotesize \textbf{Robust detection of structural break in gas prices.} Top panel: Daily natural gas spot prices with the estimated break date (May 31, 2022) marked by the vertical line. Bottom panel: The sequence of robust t-statistics for the slope change parameter ($\delta$).}
    \label{fig:gas_break}
\end{figure}

\subsubsection*{Timing and magnitude of the shock}

Figure \ref{fig:gas_break} reports the results of the robust monitoring procedure. 
The estimated break date $\hat{\tau}$, identified as the minimiser of  Equation (\ref{eq:optimization}) falls on May 31, 2022.

This break point effectively partitions the sample into two distinct economic phases: the \textit{Inflationary Phase (2019--May 2022)}, characterized by rising fuel costs, driven by post-pandemic demand recovery and heightened geopolitical tensions, and the \textit{Correction Phase (Post-May 2022)}, where prices retreat from their highs, stabilizing at a lower mean level, albeit with continued volatility.

\subsubsection*{Link to merit order dynamics}
The timing of this structural break offers a possible explanation for the regime switching behavior observed in Section \ref{sec:regime_dynamics}. Prior to the break (2021--early 2022), the sustained rise in gas prices eroded the competitive advantage of natural gas combined cycle (NGCC) units, frequently pushing them out of the baseload and allowing coal to set the marginal price. This aligns with the ``stable blocks'' of High-MEF regimes observed in Figure \ref{fig:regime_prob}.

Conversely, the negative trend post-May 2022 restored the economic competitiveness of gas. As prices fell, gas units regained their position in the merit order. However, the market did not return to a simple stable state; rather, it entered a period of volatility where the \textit{High-MEF} (coal-driven) regime appears intermittently at the margin.

\subsection{Impact of structural break on daily MEF profiles}
\label{sec:diurnal_impact}

 To investigate how the May 2022 structural break altered the carbon intensity of the grid at the hourly level, we re-estimated the Markov Switching coefficients for two distinct sub-periods: the \textit{Inflationary phase} (2019--May 2022) and the \textit{Correction phase} (Post-May 2022).

\begin{figure}[htbp]
    \centering
    \includegraphics[width=1\linewidth]{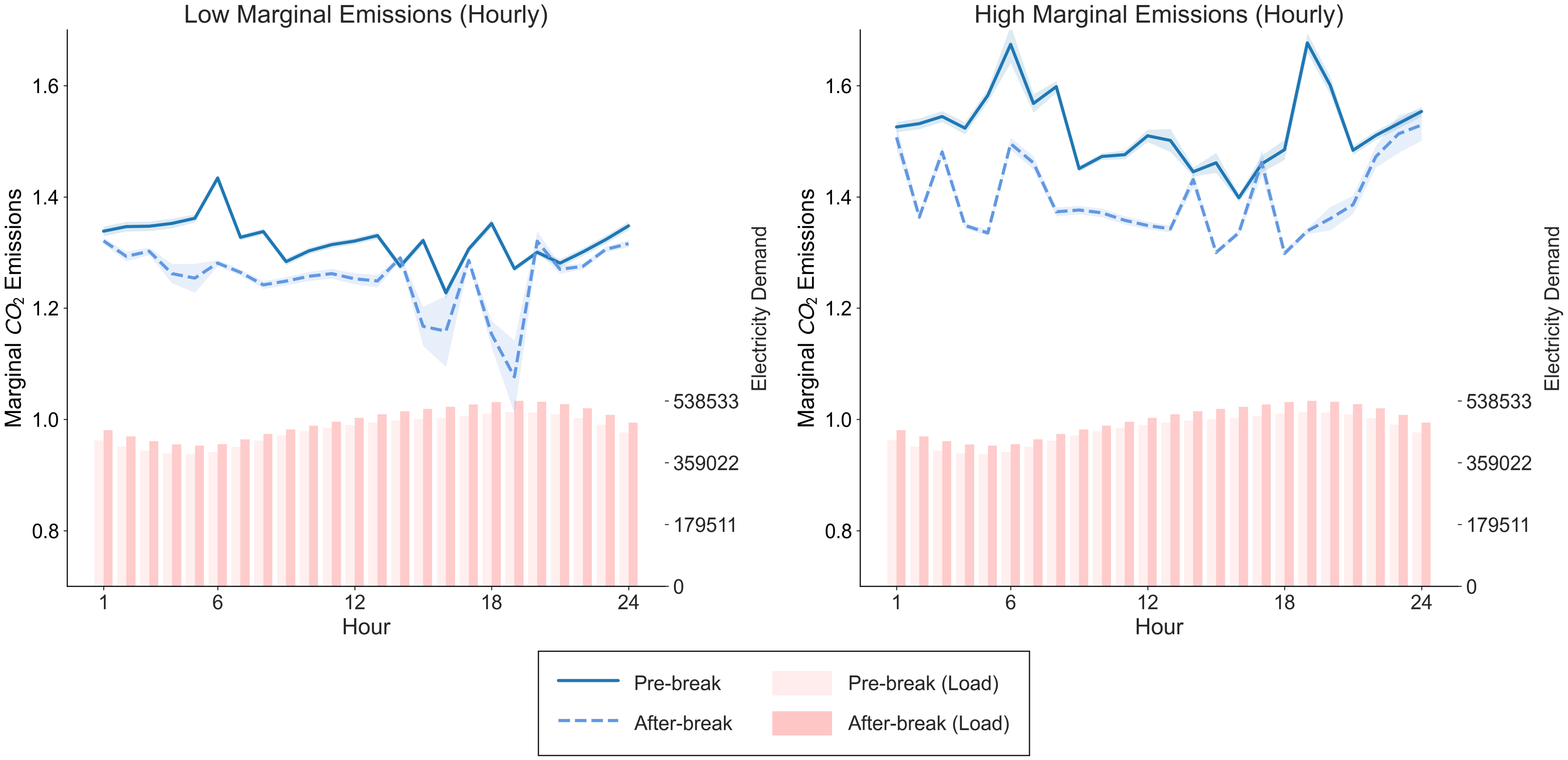} 
    \caption{\footnotesize \textbf{Structural Shift in Diurnal Marginal Emissions (2019--2025).} Comparison of hourly MEFs for the pre-break (2019--2021, solid blue) and post-break (2022--2025, dashed light blue) periods. The left panel shows the Low-Emission regime, while the right panel displays the High-Emission regime. The pink bars represent the average hourly load profile. }
    \label{fig:hourly_break_impact}
\end{figure}

Figure \ref{fig:hourly_break_impact} contrasts the hourly profiles of the \textit{Low-MEF} and \textit{High-MEF} regimes across these two phases. The results\footnote{The values for the  MEFs, SE, and average load are provided in Table S7 of the Online Supplement.} reveal a systemic reduction in marginal emissions in both states. In the \textit{Low-MEF} state (left panel), the post-break curve (dashed) lies below the pre-break baseline.
The overnight and early morning hours show a moderate decline with respect to pre-break however, during the afternoon and evening, MEFs drop sharply, reaching values as low as 1.076 at hour 19. The low regime after the break is therefore telling a story of a grid that is on average cleaner at the margin but considerably more variable, especially in hours where renewable intermittency is more pronounced.

The clearest impact of the structural break, however, is observed in the \textit{High-MEF} regime (right panel). Historically (pre-break), this regime was characterized by MEFs exceeding 1.6 lbs/kWh. Yet, in the post-break phase, emissions of this state fell by around 0.1--0.2 lbs/kWh across the day. This suggests that the decline in gas prices discussed in Section \ref{sec:structural_break} pushed coal out of the merit order even during stressed grid conditions.

\subsection{Understanding the mechanisms affecting marginal emission factors}

To disentangle the drivers of MEFs in the US48 region, we examine the evolution of the generation mix through two distinct lenses: \textit{average generation} (the volume of fuel consumed) and \textit{marginal responsiveness} (the probability that a specific fuel meets the next unit of demand). While the share of coal in baseload generation has historically declined, marginal emissions may increase or decrease depending on whether coal or gas becomes the primary marginal source (\citealp{knittel2015}, and \citealp{Holland2022}).

\subsubsection*{Econometric estimation of marginal responsiveness}
Our statistical approach for examining this structural shift adapts the methodology established by \cite{Holland2022} for estimating marginal emissions. We estimate the responsiveness of coal and gas to changes in system load using the following linear regression model for each year $y$:

\begin{equation}
    Gen_{f,t} = \beta_{f,y} \cdot Load_{t} + \alpha_{m,h} + \epsilon_{t}
    \label{eq:marginal_gen}
\end{equation}

\noindent where $Gen_{f,t}$ is the hourly aggregate generation from fuel source $f$ at hour $t$, $Load_{t}$ is the hourly system-wide electricity demand, $\alpha_{m,h}$ represents \textit{month-of-sample} by \textit{hour-of-day} fixed effects. This term absorbs predictable variation due to seasonality and daily schedules, ensuring that the estimate is driven by stochastic deviations in load rather than predictable cycles. $\beta_{f,y}$ is the coefficient of interest, representing the marginal response rate, which can be interpreted as the probability that the marginal unit of load is met by generation from fuel source $f$.

\begin{figure}[ht]
    \centering
    \includegraphics[width=1\linewidth]{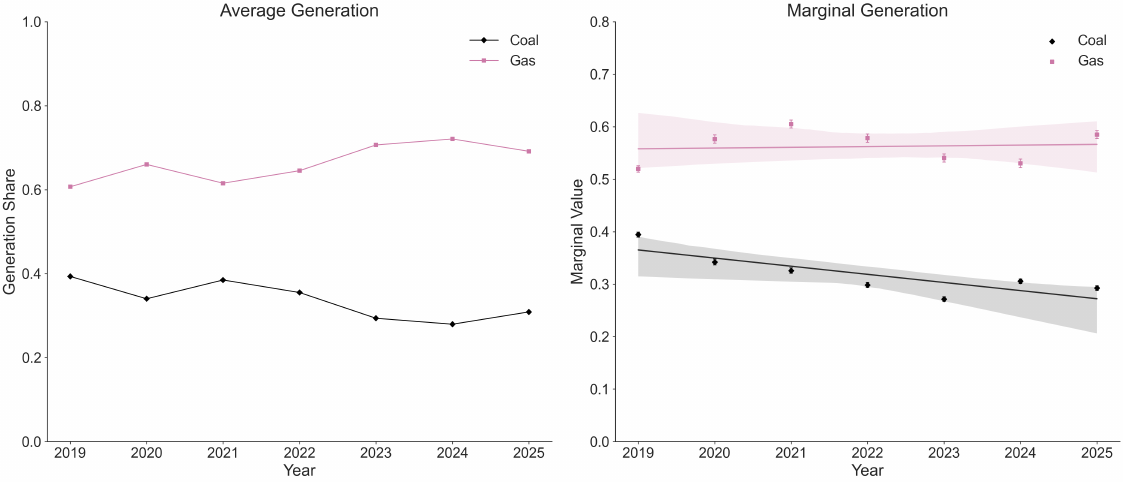}
    \caption{\footnotesize \textbf{Fuel Mix Evolution.} Comparison of average generation shares (Left) and marginal responsiveness (Right) for coal and gas. The data reveal a structural shift as coal's marginal role diminishes significantly over the considered period.}
    \label{fig:Coal_Gas}
\end{figure}

Figure \ref{fig:Coal_Gas} illustrates the evolution of the fuel mix dynamics over the considered period. The left panel displays the average generation share for coal and gas, while the right panel reports their marginal responsiveness.

\begin{figure}[h]
    \centering
    \includegraphics[width=1\linewidth]{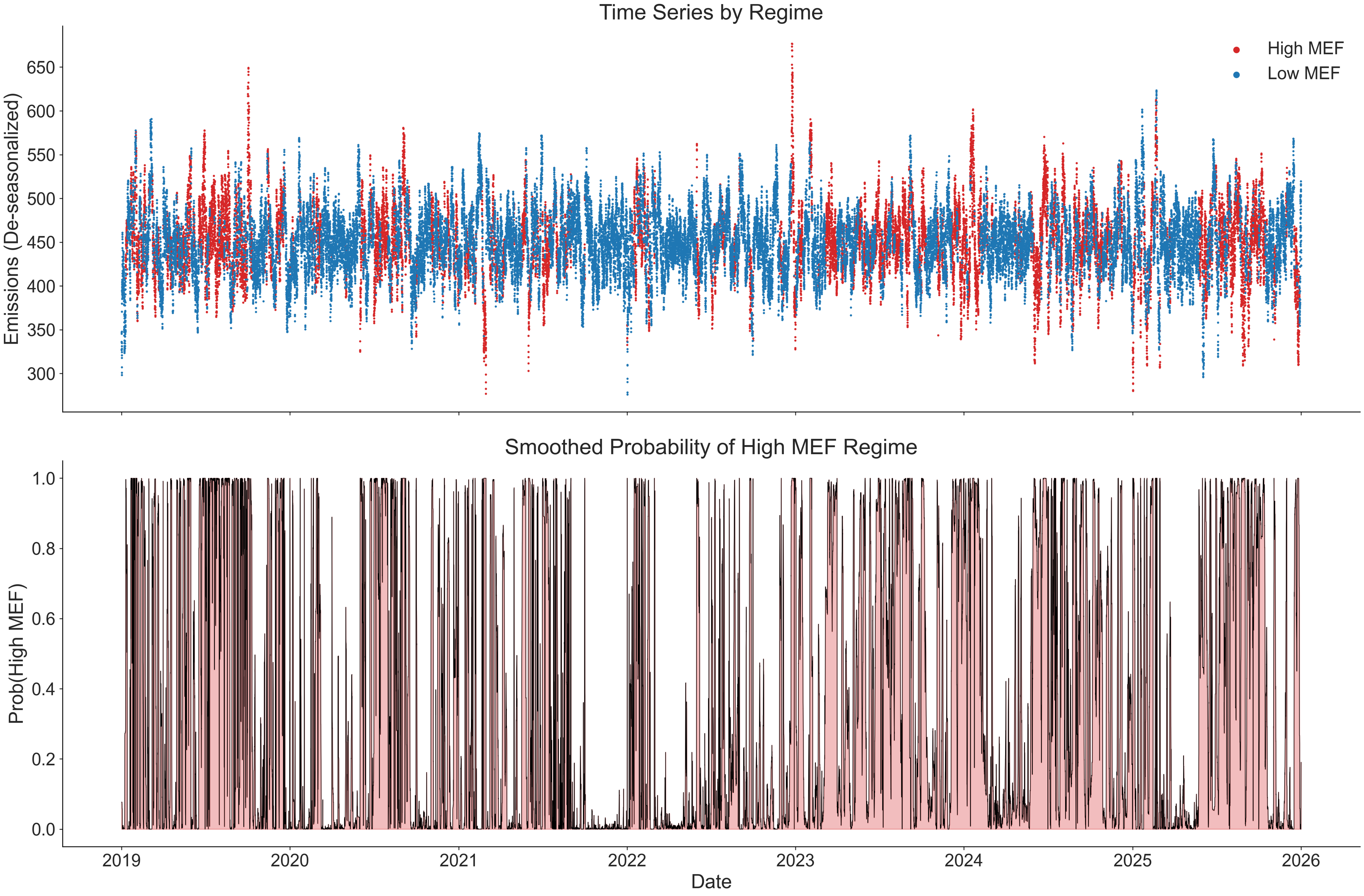} 
    \caption{\footnotesize \textbf{Regime Switching Dynamics for dummy model (2019--2025).} Top: Hourly emissions colored by the most probable regime (Red = High MEF, Blue = Low MEF). Bottom: Smoothed probability $\mathbb{P}(S_t = \text{High MEF} | \mathcal{Y}_T)$.}
    \label{fig:regime_prob_dummy}
\end{figure}

As shown in the left panel, the spread between gas and coal shares was relatively narrow in 2021. However, this gap has widened significantly in subsequent years, establishing gas as the dominant source on average. A similar divergence is observed in the right panel regarding marginal responsiveness: while the probability of gas being at margin slightly increased, the marginal role of coal has decreased sharply. The distinction between a coal-driven and gas-driven regime is therefore fading, as the system is effectively collapsing toward a single regime dominated by gas in both average and marginal emissions.

\begin{figure}[h!]
    \centering
    \includegraphics[width=0.8\textwidth]{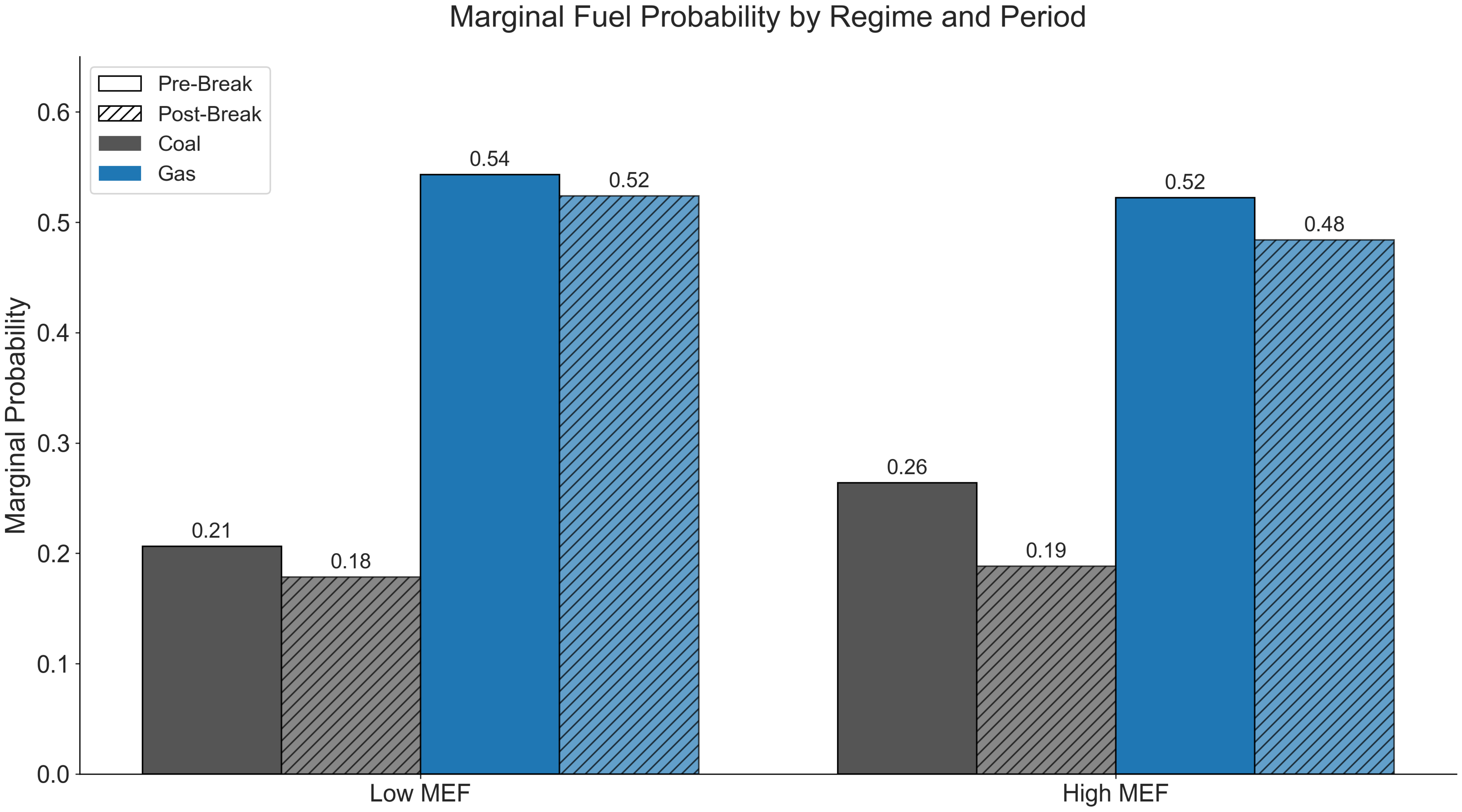}
    \caption{\footnotesize\textbf{Marginal Fuel Probability by Regime and Period.} The bar chart displays the estimated $\beta$ coefficients from Equation (\ref{eq:marginal_fuel}) for the Pre-Break (2019–May 2022) and Post-Break (Post-May 2022) periods. While Gas remains the dominant marginal source across both states, the Post-Break period shows a marked convergence where the marginal responsiveness of coal becomes nearly indistinguishable between regimes, signaling a system increasingly dominated by natural gas.}
    \label{fig:marginal_probability_pre_post_break}
\end{figure}

\subsection{MEF regimes and the energy crisis}

Finally, we investigate the impact of the structural break on marginal emission regimes using an interaction with an indicator variable equal to 1 if May 2022 or later and 0 otherwise. The appropriate Markov-switching specification can be obtained by modifying Equation(\ref{eq:observation}) in the following way:
\begin{equation}
y_t = \alpha_{s_t} +  x_t^T\beta_{s_t} +  \left(D_t \times
x_t\right)^T \delta_{s_t} + \varepsilon_t, \qquad \varepsilon_t \sim
\mathcal{N}(0,\sigma^2),
\label{eq: Dummy_MEF}
\end{equation}
where $x_t =[x_{1,t},x_{2,t}]$ and $D_t$ represents the post-2022 indicator that captures a structural change in marginal emissions.
The implied marginal emission factors can be interpreted as $\beta_{s_t}$ for the after break period and $\beta_{s_t}
+ \delta_{s_t}$ for the post-break period where $\delta_{s_t}$ accounts for the post-crisis
change in marginal emissions within each regime.

The results\footnote{ Table S8 of the Online Supplement provides the coefficients and SEs of the models estimated by Equations (\ref{eq:observation}) and (\ref{eq: Dummy_MEF}).} show that the interaction terms $\delta_{s_t}$ are negative and statistically significant in both post-break regimes and the transition probabilities in the model with the interaction terms no longer display the same pattern observed in the baseline model without the event indicator. This transition induced lasting changes in the generation mix and
dispatch order. Even as gas prices declined, the marginal emissions response to load remained
structurally altered relative to the pre-crisis period. These features clearly feature in the regime dynamics and marginal fuel probabilities in Figures  \ref{fig:regime_prob_dummy} and \ref{fig:marginal_probability_pre_post_break}, respectively.

\section{Conclusion}
\label{sec: Conclusion}

This study uncovers the existence of MEF regimes in the U.S. power sector. By employing MS-ARX, we show that canonical linear models hide important dynamics, notably the evolution of distinct ``low'' and ``high'' fossil-fuel regimes and their apparent convergence over time. We augment the analysis by identifying a structural break at the height of the global energy crisis (May 2022). Accounting for this gas price shock reveals a common structural shift within existing regimes rather than the emergence of a new state.

Our approach is also novel in that, upon controlling for multiple regimes and structural breaks, we are able to explain why MEFs are now trending \textit{downward}, counter to the increasing trend in 2010--2019 reported by \cite{Holland2022}. These findings suggest that executive orders designed to subvert the scheduled retirement of coal-fired plants to meet growing demand may only serve to exert upward pressure on both emissions and prices at the margin.

\section*{Data Availability Statement}

The data supporting the findings of this study are publicly available. Hourly electricity generation and emissions data were sourced from \citet{EIA2026}, and natural gas pricing data were obtained from the EIA Henry Hub spot price index (\citealp{EIA_HenryHub}). To ensure scientific transparency and reproducibility, the complete processing pipeline, model implementation, and source code are available at the following GitHub repository: \href{https://github.com/APanico12/STATE_DEPENDET_MEF.git}{STATE\_DEPENDET\_MEF}.

\newpage
\bibliographystyle{apalike}
\bibliography{references}

\newpage
\appendix

\setcounter{section}{0}
\setcounter{table}{0}
\setcounter{figure}{0}
\renewcommand{\thesection}{S\arabic{section}}
\renewcommand{\thetable}{S\arabic{table}}
\renewcommand{\thefigure}{S\arabic{figure}}

\centerline{\Large \textbf{Supplementary Material}}
\vspace{1cm}
\section{Summary Statistics}

\begin{table}[h!]
\centering
\caption{\textbf{Summary Statistics by Season.} This table reports the mean, standard deviation, minimum, and maximum values for Hourly Emissions (lbs) and Generation (MWh) across different seasons from 2019 to 2025.}
\label{tab:seasonal_summary}
\footnotesize
\begin{adjustbox}{width=\textwidth}
\begin{tabular}{lrrrrrrrr}
\toprule
& \multicolumn{4}{c}{\textbf{Hourly Emissions (lbs)}} & \multicolumn{4}{c}{\textbf{Hourly Generation (MWh)}} \\
\cmidrule(lr){2-5} \cmidrule(lr){6-9}
\textbf{Season} & \textbf{Mean} & \textbf{Std} & \textbf{Min} & \textbf{Max} & \textbf{Mean} & \textbf{Std} & \textbf{Min} & \textbf{Max} \\
\midrule
\multicolumn{9}{l}{\textit{Spring}} \\
2019 & 153,120 & 28,093 & 87,727 & 243,056 & 415,025 & 49,534 & 312,704 & 566,117 \\
2020 & 124,767 & 21,065 & 80,496 & 199,931 & 383,655 & 41,681 & 296,842 & 532,255 \\
2021 & 139,810 & 21,836 & 92,283 & 221,124 & 401,748 & 43,360 & 312,529 & 575,181 \\
2022 & 144,154 & 20,899 & 97,387 & 232,358 & 422,801 & 46,712 & 329,023 & 627,329 \\
2023 & 133,167 & 19,278 & 89,607 & 192,238 & 418,630 & 40,764 & 328,878 & 558,459 \\
2024 & 128,306 & 19,514 & 89,758 & 193,940 & 430,736 & 45,697 & 341,008 & 591,933 \\
2025 & 133,520 & 17,951 & 89,692 & 189,301 & 437,081 & 42,303 & 347,238 & 579,171 \\
\addlinespace
\multicolumn{9}{l}{\textit{Summer}} \\
2019 & 208,306 & 47,045 & 107,182 & 304,093 & 510,028 & 86,082 & 345,647 & 704,164 \\
2020 & 202,905 & 48,252 & 93,482 & 300,184 & 507,714 & 88,469 & 324,232 & 687,997 \\
2021 & 219,163 & 39,930 & 122,107 & 305,224 & 519,315 & 84,681 & 333,228 & 715,022 \\
2022 & 210,816 & 39,168 & 115,397 & 299,069 & 533,204 & 86,482 & 347,843 & 735,343 \\
2023 & 200,827 & 40,388 & 111,557 & 291,720 & 531,102 & 86,576 & 350,066 & 734,735 \\
2024 & 200,108 & 39,262 & 107,060 & 292,553 & 547,909 & 83,441 & 362,731 & 737,011 \\
2025 & 202,693 & 36,899 & 112,535 & 292,512 & 554,873 & 87,809 & 360,166 & 755,638 \\
\addlinespace
\multicolumn{9}{l}{\textit{Autumn}} \\
2019 & 173,055 & 36,348 & 101,654 & 280,125 & 438,322 & 66,274 & 315,900 & 653,035 \\
2020 & 154,824 & 30,538 & 97,085 & 265,489 & 415,751 & 56,014 & 317,800 & 633,456 \\
2021 & 161,059 & 30,573 & 103,764 & 261,606 & 431,537 & 58,133 & 317,510 & 624,411 \\
2022 & 157,974 & 30,052 & 90,132 & 263,658 & 435,311 & 60,365 & 266,277 & 642,757 \\
2023 & 155,580 & 29,078 & 92,352 & 273,187 & 444,760 & 61,597 & 296,158 & 697,800 \\
2024 & 151,882 & 26,538 & 101,477 & 234,034 & 449,898 & 56,112 & 327,225 & 622,791 \\
2025 & 159,825 & 25,947 & 94,952 & 242,684 & 463,077 & 58,036 & 310,793 & 629,455 \\
\addlinespace
\multicolumn{9}{l}{\textit{Winter}} \\
2019/20 & 153,481 & 22,742 & 98,422 & 221,682 & 424,921 & 37,425 & 334,553 & 542,145 \\
2020/21 & 180,941 & 26,648 & 112,598 & 252,382 & 454,154 & 38,307 & 343,078 & 553,033 \\
2021/22 & 170,805 & 29,296 & 101,161 & 248,005 & 457,730 & 41,679 & 336,901 & 584,792 \\
2022/23 & 161,609 & 29,076 & 108,197 & 272,148 & 457,206 & 43,077 & 352,285 & 632,995 \\
2023/24 & 162,237 & 30,886 & 99,372 & 263,728 & 463,787 & 44,530 & 364,881 & 643,682 \\
2024/25 & 179,923 & 32,059 & 106,155 & 278,738 & 492,871 & 50,514 & 369,560 & 671,200 \\
\bottomrule
\end{tabular}
\end{adjustbox}
\end{table}

\clearpage

\section{Selection of the Number of Regimes}

To determine the optimal number of latent regimes ($k$) for the Markov Switching Model (MSM), we estimated specifications for $k=2$ and $k=3$ under $p=1$ using the full sample period (2019--2025). The model selection criteria are presented in Table \ref{tab:s1_model_fit}.

\begin{table}[h!]
    \centering
    \caption{\footnotesize\textbf{Model fit statistics for Markov Switching specifications.} Comparison of log-likelihood and information criteria for 2-state and 3-state specifications.}
    \label{tab:s1_model_fit}
    \vspace{0.2cm}
    \begin{tabular}{lcccc}
        \toprule
        \textbf{States ($k$)} & \textbf{Log-Likelihood} & \textbf{AIC} & \textbf{BIC} & \textbf{HQIC} \\ 
        \midrule
        2 & 111,455.35 & -222,888.69 & -222,789.42 & -222,857.89 \\ 
        3 & 112,419.08 & -224,800.17 & -224,628.70 & -224,746.96 \\ 
        \bottomrule
    \end{tabular}
    \par\medskip
    \begin{minipage}{\linewidth}
        \footnotesize
        \textit{Note:} AIC = Akaike Information Criterion; BIC = Bayesian Information Criterion; HQIC = Hannan-Quinn Information Criterion. All models achieved convergence. The 3-state specification provides a 0.86\% improvement in Log-Likelihood relative to the 2-state model.
    \end{minipage}
\end{table}

\noindent \textbf{Justification for Model Selection:} \\
While the information criteria (AIC, BIC, and HQIC) are minimized for the 3-state specification, statistical fit must be weighed against economic interpretability and regime persistence. A post-estimation diagnostic of the smoothed probabilities (reported in Table \ref{tab:s2_regime_freq}) revealed that the 3-state model is over-specified.

Specifically, the regime identified as `Regime 1' captured only $0.8\%$ of the sample observations. In a Markov Switching framework, a state with such negligible probability mass (frequency $<1\%$) does not represent a persistent market structure but rather captures transient anomalies or outliers.

Consequently, we rejected the 3-state model in favor of the parsimonious 2-state specification, which identifies two robust, persistent regimes that align with the physical merit order.

\begin{table}[h!]
    \centering
    \caption{\textbf{Regime Frequency Diagnostics (3-State Model).} The extremely low frequency of Regime 1 indicates model over-specification.}
    \label{tab:s2_regime_freq}
    \vspace{0.2cm}
    \begin{tabular}{lcl}
        \toprule
        \textbf{Regime} & \textbf{Frequency (\%)} & \textbf{Classification Status} \\ 
        \midrule
        Regime 0 & 74.9\% &  Valid  \\ 
        Regime 1 & \textbf{0.8\%} & \textbf{Rejected (Transient/Outlier)} \\ 
        Regime 2 & 24.3\% &  Valid  \\ 
        \bottomrule
    \end{tabular}
    \par\medskip
    \footnotesize
    \textit{Note:} Frequency is defined as the percentage of hours where the smoothed probability of the regime exceeds that of all other regimes ($P(S_t = k | I_T) > P(S_t \neq k | I_T)$).
\end{table}

\newpage

\section{Emissions Diagnostics (2019--2025)}

\makeDashboard{Emissions}{2019}
\makeDashboard{Emissions}{2020}
\makeDashboard{Emissions}{2021}
\makeDashboard{Emissions}{2022}
\makeDashboard{Emissions}{2023}
\makeDashboard{Emissions}{2024}
\makeDashboard{Emissions}{2025}

\section{Generation Diagnostics (2019--2025)}

\makeDashboard{Generation}{2019}
\makeDashboard{Generation}{2020}
\makeDashboard{Generation}{2021}
\makeDashboard{Generation}{2022}
\makeDashboard{Generation}{2023}
\makeDashboard{Generation}{2024}
\makeDashboard{Generation}{2025}

\section{Autoregressive Order Selection}

The selection of the autoregressive lag order ($p$) is a critical specification decision in regime-switching frameworks. While standard linear diagnostics (e.g., Partial Autocorrelation Functions) on the de-seasonalized series suggested potential higher-order dependence, these linear tools do not account for the non-linear dynamics introduced by regime switching.
To rigorously determine the optimal lag structure, we systematically estimated MS-ARX specifications with $p \in \{1, 2, 3, 4\}$. The results of this sensitivity analysis are summarized in Table~\ref{tab:lag_sensitivity}.
We observe a clear trade-off between model complexity and numerical stability:
 Models with $p \geq 2$ systematically failed to satisfy convergence criteria. While these models report higher log-likelihoods and more favorable information criteria (AIC/BIC), these values are numerical artifacts of an unstable optimization. This convergence failure is a well-documented identification issue in regime-switching literature (\citealp{hamilton1990}, and \citealp{Krolzig1997}) and arises from two sources: 1) higher-order lags exponentially increase the parameter space - an MS-ARX(2) model with $K=2$ regimes requires estimation of $2K(p+3)$ parameters - creating a highly non-convex likelihood surface prone to local optima; 2) autoregressive dynamics and regime persistence are observationally similar; the model cannot reliably distinguish whether emission persistence arises from AR($p$) terms or from regime stickiness (high $p_{11}$, $p_{22}$ transition probabilities).

\begin{table}[htbp]
    \centering
    \begin{threeparttable}
        \caption{Convergence Diagnostics for MS-ARX Models with Alternative Lag Structures}
        \label{tab:lag_sensitivity}
        \begin{tabular}{cccccc}
            \toprule
            $p$ & Regimes & Log-Likelihood & AIC & BIC & HQIC \\
            \midrule
            1 & 2 & 110,685.12 & $-221,348.25$ & $-221,248.98$ & $-221,317.45$ \\
            2 & 2 & 113,571.17$^{\dagger}$ & $-227,116.35$ & $-226,999.03$ & $-227,079.95$ \\
            3 & 2 & 113,591.85$^{\dagger}$ & $-227,153.69$ & $-227,018.32$ & $-227,111.69$ \\
            4 & 2 & 113,644.60$^{\dagger}$ & $-227,255.20$ & $-227,101.79$ & $-227,207.60$ \\
            \bottomrule
        \end{tabular}
        \begin{tablenotes}
            \small
            \item \textit{Note:} All models were estimated with homoscedastic errors ($\sigma_1^2 = \sigma_2^2$).
            \item $^{\dagger}$ Likelihood values for non-convergent models. These values reflect local maxima or saddle points rather than global optima.
        \end{tablenotes}
    \end{threeparttable}
\end{table}

Consequently, the MS-ARX(1) specification is the highest-order model that achieves stable convergence. This selection is also theoretically consistent with power system dynamics: the AR(1) term captures the short-run physical inertia of the grid (e.g., thermal ramping constraints), while the Markov process absorbs the structural shifts in the merit order (e.g., the discrete switch between gas-margin and coal-margin states). We therefore proceed with the parsimonious MS-ARX(1) specification for the main analysis.

\section{Residuals Diagnostics (2019--2025)}

\makeDashboard{MSM}{2019}
\makeDashboard{MSM}{2020}
\makeDashboard{MSM}{2021}
\makeDashboard{MSM}{2022}
\makeDashboard{MSM}{2023}
\makeDashboard{MSM}{2024}
\makeDashboard{MSM}{2025}

\section{Comparative Analysis of Model Fit}

To validate the selection of the Markov Switching Autoregressive model (MS-ARX), we compared its goodness-of-fit against two benchmark specifications: US-FE approach (\citealp{Holland2008}) and  ARIMA model (\citealp{Beltrami2020}). To ensure the comparability of log-likelihood functions across different model classes, all specifications were fitted to the standardized series of hourly marginal emissions.

Table \ref{tab:model_comparison} summarizes the performance metrics. First, the results highlight that the static OLS framework lacksacks in explaining the data behavior, noting the wide gap between MSM-ARX(1) and ARIMA.
The comparison between the dynamic specifications highlights a fundamental trade-off between statistical curve-fitting and economic structure.

The ARIMA(3,0,4) model achieves the highest statistical fit relying on a complex lag structure with seven temporal parameters. Although mathematically flexible, this specification imposes a restrictive assumption: it treats the data-generating process as a single, invariant linear mechanism, obscuring the physical reality of fuel switching.
    
The MS-ARX(1) model, employing a parsimonious first-order lag structure ($p=1$), achieves a competitive fit with respect to ARIMA while explicitly modeling structural non-linearities. Although the ARIMA model offers a marginally superior statistical fit, its predictions average over distinct market conditions, leading the model to not have a useful economic specification.

\begin{table}[htbp]
    \centering
    \begin{threeparttable}
        \caption{Goodness-of-Fit Comparison}
        \label{tab:model_comparison}
        \begin{tabular}{lcccc}
            \toprule
            Model Specification & Log-Likelihood & AIC & BIC & HQIC \\
            \midrule
            ARIMA(3,0,4)  & 113,855.27 & $-227,688.55$ & $-227,589.28$ & $-227,657.75$ \\
            MS-ARX(1)     & 110,685.12 & $-221,348.25$ & $-221,248.98$ & $-221,317.45$ \\
            OLS (US-FE)   & 21,690.97  & $-43,375.94$  & $-43,348.87$  & $-43,367.54$ \\
            \bottomrule
        \end{tabular}
        \begin{tablenotes}
            \small
           
            \item \textit{Note:} All models were fitted to the standardized series of hourly marginal emissions to ensure scale comparability of the log-likelihood functions.
        \end{tablenotes}
    \end{threeparttable}
\end{table}

\section{Annual Model Estimation Results}

Table \ref{tab:annual_mef_estimates} details the estimated Marginal Emission Factors (MEFs) for each year from 2019 to 2025 across the different methodological approaches. The values represent the marginal CO$_2$ emissions (in lbs/MWh) induced by a unit increase in electricity demand. Standard errors are provided in parentheses.

We observe a consistent ordering among the benchmark models: the static Fixed Effects model (US-FE) consistently yields the highest MEF estimates (ranging from 1.37 to 1.50 lbs/MWh), while the dynamic approaches (Hawkes and ARIMA) provide more conservative estimates.

The Markov Switching results reveal significant heterogeneity hidden by the linear benchmarks. The \textit{High-MEF} regime estimates remain robustly high ($1.25 - 1.55$ lbs/MWh) across all years, effectively capturing the "coal-on-margin" dynamics. In contrast, the \textit{Low-MEF} regime shows substantial volatility. Notably, in 2019, the \textit{Low-MEF} regime was 0.465 lbs/MWh, indicative of a gas-dominated margin. However, during the gas price inflationary period (2021-2022), the \textit{Low-MEF} regime converged closer to the \textit{High-MEF} regime (e.g., 1.28 lbs/MWh in 2021), reflecting the economic displacement of coal into the baseload and the increased marginality of less efficient units.

\begin{table}[ht!]
\centering
\caption{Annual Marginal Emission Factors (lbs/MWh) by Model. Standard Errors are reported in parentheses.}
\label{tab:annual_mef_estimates}
\resizebox{\textwidth}{!}{%
\begin{tabular}{l c c c c c}
\hline
\textbf{Year} & \textbf{US-FE (OLS)} & \textbf{Hawkes (FD)} & \textbf{ARIMA} & \textbf{MS-High} & \textbf{MS-Low} \\
\hline
2019 & 1.500 (0.016) & 1.372 (0.002) & 1.397 (0.002) & 1.500 (0.004) & 0.465 (0.113) \\
2020 & 1.415 (0.018) & 1.327 (0.002) & 1.368 (0.001) & 1.412 (0.003) & 0.916 (0.081) \\
2021 & 1.395 (0.017) & 1.244 (0.002) & 1.315 (0.001) & 1.546 (0.014) & 1.281 (0.003) \\
2022 & 1.390 (0.018) & 1.219 (0.002) & 1.275 (0.001) & 1.255 (0.003) & 0.891 (0.018) \\
2023 & 1.369 (0.017) & 1.283 (0.001) & 1.292 (0.001) & 1.285 (0.003) & 0.877 (0.044) \\
2024 & 1.415 (0.016) & 1.292 (0.002) & 1.311 (0.001) & 1.341 (0.008) & 1.297 (0.003) \\
2025 & 1.392 (0.012) & 1.232 (0.002) & 1.260 (0.001) & 1.393 (0.004) & 1.236 (0.009) \\
\hline
\end{tabular}}
\end{table}

\section{Hourly Marginal Emission Factor (MEF) Results}

The hourly Marginal Emission Factor (MEF) estimates for both operational periods are summarized in Table \ref{tab:mef_summary}. The coefficients $\beta_{\text{low}}$ and $\beta_{\text{high}}$ represent the marginal impact of load on emissions, accompanied by their respective standard errors ($SE$) and the average system load ($\bar{L}$).
\newpage

\begin{table}[H]
\centering
\small
\caption{Hourly MEF Summary: Pre-break and After-break Periods}
\label{tab:mef_summary}
\begin{tabular}{l S[table-format=1.3] S[table-format=1.3] S[table-format=1.4] S[table-format=1.4] S[table-format=6.0]}
\toprule
{Hour ($h$)} & {$\beta_{\text{low}}$} & {$\beta_{\text{high}}$} & {$SE_{\text{low}}$} & {$SE_{\text{high}}$} & {Avg. Load ($\bar{L}$)} \\
\midrule
\multicolumn{6}{c}{\textit{Panel A: Pre-break Period}} \\
\midrule
1  & 1.338 & 1.526 & 0.0085 & 0.0092 & 424110 \\
2  & 1.346 & 1.532 & 0.0087 & 0.0096 & 406299 \\
3  & 1.347 & 1.544 & 0.0085 & 0.0098 & 393565 \\
4  & 1.352 & 1.523 & 0.0086 & 0.0108 & 385951 \\
5  & 1.361 & 1.582 & 0.0070 & 0.0117 & 383879 \\
6  & 1.434 & 1.674 & 0.0062 & 0.0338 & 389642 \\
7  & 1.327 & 1.568 & 0.0051 & 0.0166 & 405185 \\
8  & 1.338 & 1.598 & 0.0051 & 0.0102 & 423206 \\
9  & 1.284 & 1.451 & 0.0063 & 0.0053 & 437960 \\
10 & 1.303 & 1.473 & 0.0059 & 0.0061 & 450008 \\
11 & 1.314 & 1.476 & 0.0059 & 0.0079 & 460178 \\
12 & 1.320 & 1.510 & 0.0056 & 0.0110 & 468584 \\
13 & 1.330 & 1.501 & 0.0061 & 0.0206 & 475399 \\
14 & 1.275 & 1.445 & 0.0072 & 0.0082 & 481029 \\
15 & 1.322 & 1.461 & 0.0062 & 0.0178 & 485505 \\
16 & 1.228 & 1.398 & 0.0097 & 0.0069 & 489412 \\
17 & 1.306 & 1.459 & 0.0063 & 0.0153 & 494437 \\
18 & 1.352 & 1.485 & 0.0075 & 0.0180 & 501134 \\
19 & 1.271 & 1.677 & 0.0051 & 0.0173 & 505446 \\
20 & 1.300 & 1.600 & 0.0051 & 0.0159 & 504602 \\
21 & 1.281 & 1.484 & 0.0072 & 0.0078 & 499789 \\
22 & 1.301 & 1.510 & 0.0072 & 0.0083 & 488481 \\
23 & 1.323 & 1.532 & 0.0071 & 0.0084 & 469167 \\
24 & 1.348 & 1.553 & 0.0075 & 0.0091 & 446454 \\
\midrule
\multicolumn{6}{c}{\textit{Panel B: After-break Period}} \\
\midrule
1  & 1.321 & 1.507 & 0.0056 & 0.0222 & 453236 \\
2  & 1.293 & 1.364 & 0.0085 & 0.0065 & 435206 \\
3  & 1.302 & 1.481 & 0.0064 & 0.0111 & 421263 \\
4  & 1.262 & 1.349 & 0.0173 & 0.0054 & 412064 \\
5  & 1.254 & 1.335 & 0.0259 & 0.0051 & 408762 \\
6  & 1.281 & 1.496 & 0.0054 & 0.0109 & 412993 \\
7  & 1.264 & 1.461 & 0.0055 & 0.0120 & 426562 \\
8  & 1.242 & 1.373 & 0.0070 & 0.0081 & 442358 \\
9  & 1.249 & 1.376 & 0.0074 & 0.0075 & 455219 \\
10 & 1.257 & 1.371 & 0.0082 & 0.0078 & 466616 \\
11 & 1.262 & 1.358 & 0.0086 & 0.0074 & 478028 \\
12 & 1.253 & 1.348 & 0.0099 & 0.0066 & 489055 \\
13 & 1.249 & 1.342 & 0.0111 & 0.0061 & 499262 \\
14 & 1.290 & 1.431 & 0.0046 & 0.0145 & 507972 \\
15 & 1.167 & 1.300 & 0.0348 & 0.0047 & 515009 \\
16 & 1.158 & 1.336 & 0.0641 & 0.0049 & 521204 \\
17 & 1.286 & 1.462 & 0.0051 & 0.0231 & 528022 \\
18 & 1.153 & 1.298 & 0.0241 & 0.0052 & 535363 \\
19 & 1.076 & 1.339 & 0.0653 & 0.0053 & 538533 \\
20 & 1.320 & 1.361 & 0.0181 & 0.0210 & 535941 \\
21 & 1.270 & 1.386 & 0.0079 & 0.0163 & 529127 \\
22 & 1.275 & 1.472 & 0.0050 & 0.0194 & 517169 \\
23 & 1.306 & 1.513 & 0.0052 & 0.0341 & 497975 \\
24 & 1.316 & 1.529 & 0.0060 & 0.0279 & 475369 \\
\bottomrule
\end{tabular}
\end{table}

\section{Dummy MEF}

Table \ref{tab:model_comparison_dummy} presents the estimated coefficients for the Markov-switching specifications defined in Equations (5) and (19). All parameters are statistically significant at the 1\% level. Notably, the interaction terms ($\delta_{s_t}$) are negative across both regimes, supporting our hypothesis that the structural break in natural gas prices following May 2022 has been beneficial in reducing marginal emissions.

\begin{table}[htbp]
\centering
\caption{Comparison of MSM Model Coefficients}
\label{tab:model_comparison_dummy}
\begin{tabular}{lcc}
\hline 
Variable & Model 1 (With Dummy) & Model 2 (Baseline) \\ \hline
 Low MEF & 1.2707 & 1.2539 \\
 & (0.0026) & (0.0016) \\
High MEF & 1.4544 & 1.4361 \\
 & (0.0036) & (0.0039) \\
Dummy\_Low & -0.1005 & -- \\
 & (0.0084) & \\
Dummy\_High & -0.0565 & -- \\
 & (0.0068) & \\ \hline 
\multicolumn{3}{l}{\small \textit{Note: Standard errors are reported in parentheses.}}    \\
\multicolumn{3}{l}{\small \textit{Note: All coefficients are statistically significant at the 1\% level.}}

\end{tabular}
\end{table}

\end{document}